\documentclass[prb,twocolumn,floatfix,amsfonts]{revtex4-1}
\usepackage{graphicx,graphics,color,epsfig}% Include figure files
\usepackage{bm}
\usepackage{amsmath}
\usepackage{amssymb}
\usepackage{epstopdf}
\usepackage{epsfig}
\usepackage{graphicx}

\usepackage{longtable}
\usepackage{rotating}
\usepackage{multirow}

\begin{document}

\title{Universal co-existence of photovoltaics and ferroelectricity from two-dimensional 3R bilayer BX (X=P, As, Sb)}

\author{Qiang Wang,\textit{$^{a,b,c}$} Yan Liang,\textit{$^{d}$} Hui Yao,\textit{$^{a}$} Jianwei Li,\textit{$^{a}$} Tianwei Liu,\textit{$^{a}$} and Bin Wang,$^{\ast}$\textit{$^{a}$}}
\address{$^a$ ~Shenzhen Key Laboratory of Advanced Thin Films and Applications, College of Physics and Optoelectronic Engineering, Shenzhen University, Shenzhen, 518060, People's Republic of China. E-mail: binwang@szu.edu.cn}
\address{$^b$ ~Key Laboratory of Optoelectronic Devices and Systems of Ministry of Education and Guangdong Province, College of Optoelectronic Engineering, Shenzhen University, Shenzhen 518060, China.}
\address{$^c$ ~School of Materials Science and Engineering, Hanshan Normal University, Chaozhou 521041, People's Republic of China.}
\address{$^d$ ~School of Physics, State Key Laboratory of Crystal Materials, Shandong University, 250100 Jinan, People's Republic of China.}

\begin{abstract}

The intertwined ferroelectricity and photovoltaics in two-dimensional (2D) materials will enable the favorable improvement and control of photovoltaic preformances. In this paper, we take 2D 3R bilayer BX (X=P, As, Sb) as model systems to study the photovoltaic characteristics of intrinsic 2D out-of-plane (OOP) ferroelectric material, and try to explore a strategy to regulate the photoelectric properties by changing the strength of ferroelectric polarization. Due to the spatial inversion symmetry broken caused by special 3R stacking, spontaneous OOP ferroelectric polarization will appear in the 3R bilayer BX, which can be swichable through a specific interlayer sliding. The OOP ferroelectricity leads to charge transfer between layers, realizes efficient spatial separation between holes and electrons, and forms the characteristics of type-II band alignment. Moreover, due to perfect lattice match on account of two identical layers, the 3R bilayer BX is more stable and easy to realize in experiments than most of traditional 2D heterostructures made up of different materials. The 3R bilayer BX shows moderate band gap, ultra-high carrier mobility and efficient optical absorption, and its nano-devices present large photocurrent, high photon responsivity and excellent external quantum efficiency. More importantly, all these photoelectric parameters depend on the intrinsic OOP ferroelectric strength. By changing the interlayer distance of bilayer BX, the ferroelectric polarization can be regulated effectively to achieve the optimal photoelectric performance. Finally, we emphasize the importance and universality of spatial inversion symmetry broken in layered materials beyond the 3R bilayer BX to realize the co-regulation of ferroelectric and photovoltaics.
\end{abstract}

\maketitle

\section{Introduction}

The reversible spontaneous polarization of ferroelectric materials enable them desirable for high-density data storage and fast device operation without quantum tunneling or high power consumption/dissipation.\cite{OOP1, OOP2} Traditional ferroelectric materials were commonly based on some complex oxides including ABO$_{3}$ perovskites\cite{OOP3}, lithium niobate (LiNbO$_{3}$) ferroelectric crystal\cite{AM26,AM30}, BaBi$_{2}$Ta$_{2}$O$_{9}$\cite{OOP4}, and plenty of their investigations and commercial applications have been proposed in the last few decades\cite{OOP5,OOP6}. With continue miniaturization of electronic devices and rapid development of 2D materials, more efforts have been made toward exploring ferroelectricity in 2D thin film materials.\cite{OOP7}. More strikingly, the out-of-plane (OOP) ferroelectricity in these 2D thin films is more favorable than the in-plane one because the vertical polarizability is more desirable for most practical technologies\cite{ACS3,ACS4}. The most direct idea is to process the bulk ferroelectric materials to thin film by experimental methods. However, due to the depolarizing electrostatic field and electron reconstruction, the OOP ferroelectricity will vanish in experiments when the thickness of the film is below several to tens of nanometers\cite{OOP10,OOP11}. Therefore, it is reasonable to explore 2D ferroelectric materials with intrinsic OOP polarization. So far, a series of such materials have been proposed including In$_{2}$Se$_{3}$\cite{OOP16}, MoTe$_{2}$\cite{OOP14}, WTe$_{2}$\cite{OOP15}, Sc$_{2}$CO$_{2}$\cite{OOP18}, and even the ferroelectric property has been found in 2D elemental materials, such as bilayer phosphorene, arsenene, and antimonene.\cite{OOP}  To data, exploring 2D thin films with excellent OOP ferroelectricity is still in its infancy, and hence of great significance to clarify the microscopic origin inside and further assess its fascinating facilitation in the modern micro-electronics industry.

The utilization of solar energy by photovoltaic conversion is an effective and promising method to meet the urgent requirements of energy crisis. In view of different generation process of electron-hole pairs, the photovoltaic systems can be generally classified into two groups. One is the traditional bulk inorganic semiconductors such as Si, CdTe, and GaAs,\cite{LYJMCA4} where the photo-excited electrons and holes are generated without intermediate steps but also the missing of long-range Coulomb interactions. The other refers to the excitonic solar cells (XSCs), in which the electron-hole pairs are generated upon illumination.\cite{LYJMCA5,LYJMCA6} As indicated in previous studies, the different ionization potentials and electron affinity at the donor-acceptor interfaces of XSCs would induce a high yield dissociation of excitons, and hence a substantial power conversion efficiency (PCE).\cite{LYJMCA7} However, the PCE of most recent XSCs are still below 12\% although considerable efforts have been dedicated to improve their relevant performances.\cite{LYJMCA9,LYJMCA11} Besides, more critical factors are non-negligible for high efficient XSCs, including high visible light optical absorption\cite{LYJMCA12}, carrier mobility,\cite{LYJMCA13,WQJMCA} small exciton binding energy,\cite{ZPJMCA13} and moderate direct band gap (1.2-1.6 eV) of the donor.\cite{WQKAS18,WQKAS1}

Very recently, the staggered band offset of type-II heterostructures has been confirmed to be favorable for the effective separation of the excitons, implying potential applications in the XSCs.\cite{JMCC23,WQJMCA} However, the type-II heterostructures can not be always guaranteed due to the different chemical potentials between different components (there exist three different types of semiconductor heterostructures).\cite{WQJMCC} In addition, most of the proposed type-II heterostructures are composed of different monolayer materials which are probably instability in experimental manipulation due to lattice mismatch and large potential barrier between two components. Fortunately, this problem can just be overcome in recently proposed 2D bilayer ferroelectric films because they are stacked by the same layered materials.\cite{MoS2,MoS21} Due to special stacking mode, the spatial reflection symmetry is broken, and thus a spontaneous vertical polarization between two layers is induced.\cite{VS2,LLYY} This intrinsic electrostatic field is an exciting realm in designing 2D photovoltaic devices because the electrons and holes can be separated effectively in space. Moreover, recent experimental technique of growing high-quality atomic crystals layer by layer also guarantees the feasibility of stacking individual thin film into different layered materials with van der Waals (vdWH) interaction.\cite{BN,MoS2,MoS21} In light of the charge transfer between the interfacial layers, these newly combined 2D layered materials can enrich the exceptional properties of the isolated components and break the limitations of their applications in many fields, and one of the potentially inspiring combinations is ferroelectric and photovoltaic. Therefore, it is very necessary to explore the 2D ferroelectric/photoelectric hybrid materials, where the photoelectric efficiency can be naturally improved and adjusted by the intrinsic ferroelectricity.  Moreover,  it is significant to propose a fundamental mechanism to design stable and high-efficient 2D ferroelectric/photoelectric hybrid materials  and understand their inherent physical mechanisms for diversity and efficiency.

2D hexagonal group-VA boride BX (X=P, As, Sb), derivatives of h-BN, is one of the most promising families of 2D materials, which has been successfully fabricated experimentally and widely studied in recent years.\cite{BX} Plenty of prominent physical properties have been discovered in this family, including ultra-high carrier mobilities, moderate direct band gaps and novel electronic, thermal and optical properties\emph{et.al.}. \emph{ect.}\cite{BX,WQJMCA} In this work, we take the special 3R stacking configurations of bilayer BX (X=P, As, and Sb) as examples to explore the coexist of OOP ferroelectricity and photovoltaics by using the first principles calculation method. Due to the charge redistribution induced by the unique lattice stacking, sizeable and switchable OOP ferroelectric polarizations could be detected in these crystals accompany with intrinsic type-II band alignments between two semiconducting layers. Moderate quasi-direct band gaps, ultra-high carrier mobilities and efficient optical absorptions indicate appealing applications of these materials in the stable and novel 2D XSCs. The photovoltaic performance can be adjusted effectively by changing the strength of OOP ferroelectric polarization. The decrease of OOP dipole moment can enhance PCE efficiently, while its increase can induce an obvious red-shift of optical absorption in the visible light region. Moreover, extremely large photon responsivities and external quantum efficiencies are obtained in the 3R BP based few-layer optoelectronic devices, which can be further enhanced by adjusting the OOP dipole moment. Our numerical results show that 2D OOP ferroelectricity can promote the improvement of photoelectric performance of 3R bilayer BX materials omnibearingly. More importantly, the fundamental mechanisms of all above conclusions can also be extended into other layered photovoltaic ferroelectrics, and hence all these fascinating findings predicted in 3R bilayer BX systems may shed a new guidance in future experimental design and practical application of 2D photoelectric conversion devices and XSCs.

\begin{figure*}[!tb]
\includegraphics[width=15.5cm]{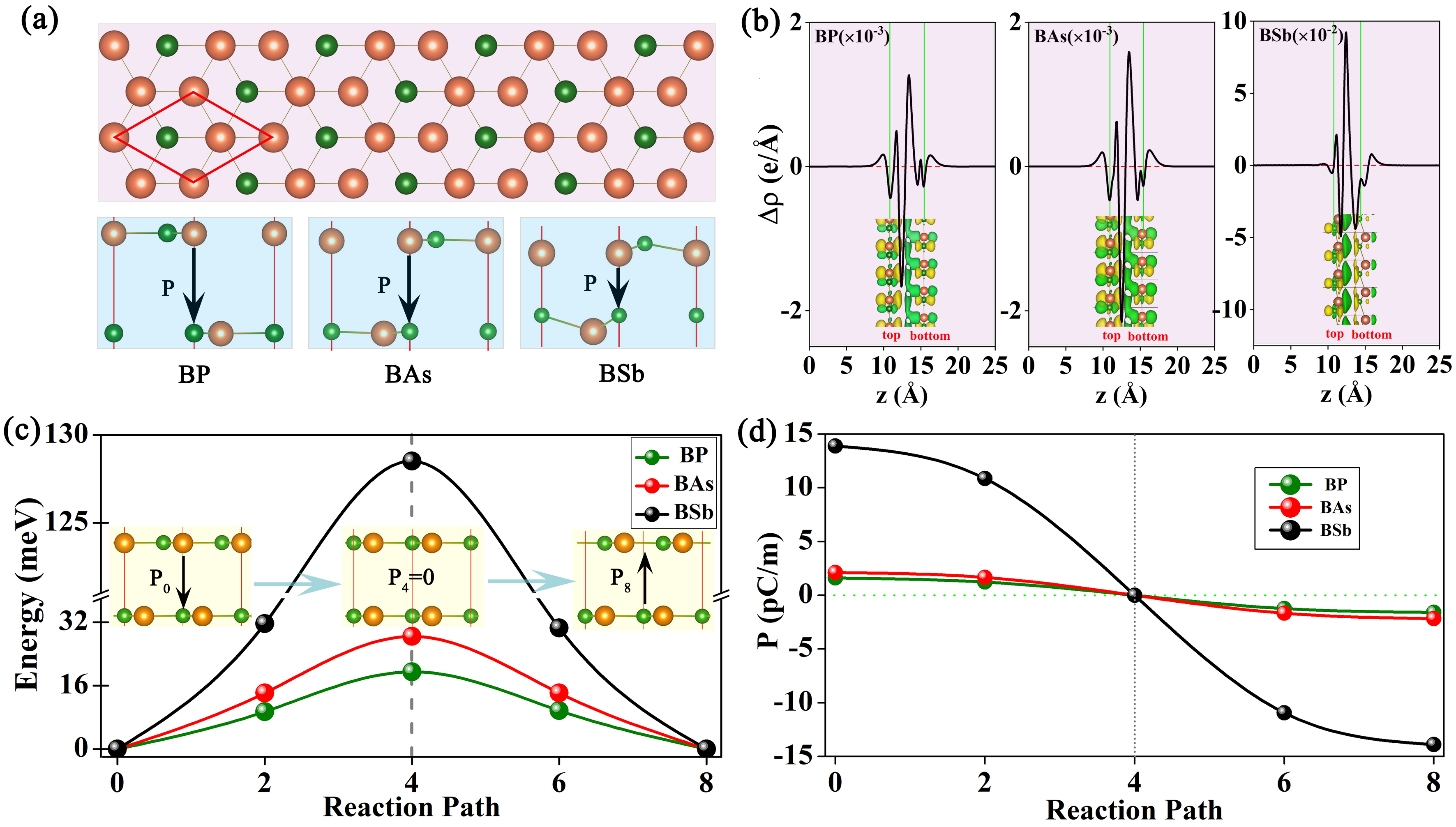}
\centering
\caption{ (color online) (a) Top and side views of schematic structures of bilayer 3R BX (X=P, As, Sb), where X and B atoms are labeled by orange and green balls. The black arrow denotes the direction of charge transfer and ferroelectric polarization. The primitive cell is marked by the rhombus. (b) Charge density deformation of plane projection and three-dimensional integration for the 3R bilayer BP, BAs and BSb. In each panel, the red horizontal dotted lines denoted the Fermi level, the green vertical lines indicate edges of the top and bottom layers, and the yellow and cyan isosurfaces represent the charge accumulation and depletion at the interface with an isovalue of 0.001 e{\AA}$^{-3}$ (BP and BAs) and 0.045 e{\AA}$^{-3}$ (BSb). (c) The reaction energy barriers and (d) the magnitudes of dipole moment along the ferroelectric switching pathway. The three inserts in (c) are side view atomic structures at initial, middle and final states in the ferroelectric switching pathway.}
\label{fig1}
\end{figure*}

The rest parts of this paper are organized as follows. In Section 2, the computational methods used in this work are briefly introduced. In Section 3, the numerical results accompany with physical mechanisms are presented, including experimental feasibility, electric polarization, ferroelectric switching and tunable electronic/photovoltaic performances of the 3R bilayer BX structures. The tunable photocurrent and photovoltaic revelent quality factors in the 3R BP based nano-devices are also performed. In section 4, a brief summary of this work is presented.

\section{Model and Numerical Method}

In this study, the structural and electronic properties of the 2D BX (X=P, As, Sb) lattices are implemented by the Vienna \emph{ab initio} simulation package (VASP) based on the density functional theory (DFT).\cite{JMCA40} The generalized gradient approximation (GGA) in the form of Perdew-Burke-Ernzerhof (PBE)\cite{JMCA41} and Heyd-Scuseria-Ernzerhof (HSE06)\cite{JMCA44} are used to deal with the exchange-correlation functional. The projected augmented wave (PAW) approach\cite{PAW} is employed to manipulate the ion-electron interaction. A cut-off energy of 500 eV is set for the plane wave function, and a vacuum larger than 20 {\AA} is used to eliminate the spurious interaction perpendicular to the 2D plane. $15\times15\times1$ k-meshed grid is set in the first Brillouin zone to describe the periodicity of the 2D plane. The convergence criteria of structural relaxation is defined less than $0.01 eV/${\AA} for force and $10^{-5} eV$ for energy. The phonon spectrum is implemented by the Phonopy code,\cite{KAS36} and the \emph{ab initio} molecular dynamics simulation is performed within a $3\times3\times1$ supercell at a time step of 1 fs under finite temperature. The energy barriers during ferroelectric switching are calculated by using the method of nudged elastic band (NEB),\cite{OOP34} and the ferroelectric polarization is evaluated by the Berry phase approach.\cite{OOP19,OOP23}

The photovoltaic transport in the 2D BP-based nano-device is performed by the first principles quantum transport package NanoDcal,\cite{nanodcal1,nanodcal2,KAS122} which is based on the combination of non-equilibrium Green's function (NEGF) and DFT.\cite{nanodcal1} In this calculation, the norm-conserving nonlocal pseudo-potential is used to define the atomic core\cite{KAS51}, and the wave function is expanded by the atomic orbital basis set with double-zeta polarization (DZP).  The exchange-correlation potential is handled at the PBE level.\cite{KAS52} $24\times1$ k-meshed gride is set in the self-consistent calculation, and $256\times1$ k-meshed gride is adopted in the photocurrent calculation. Finally, the convergence criteria of self-consistence is set less than $10^{-5}$ eV.

\section{Numerical Results and Discussions}

\subsection{Electric polarization and ferroelectric switching in 3R bilayer BX (X=P, As, Sb)}

\begin{figure*}[!tb]
\includegraphics[width=15.5cm]{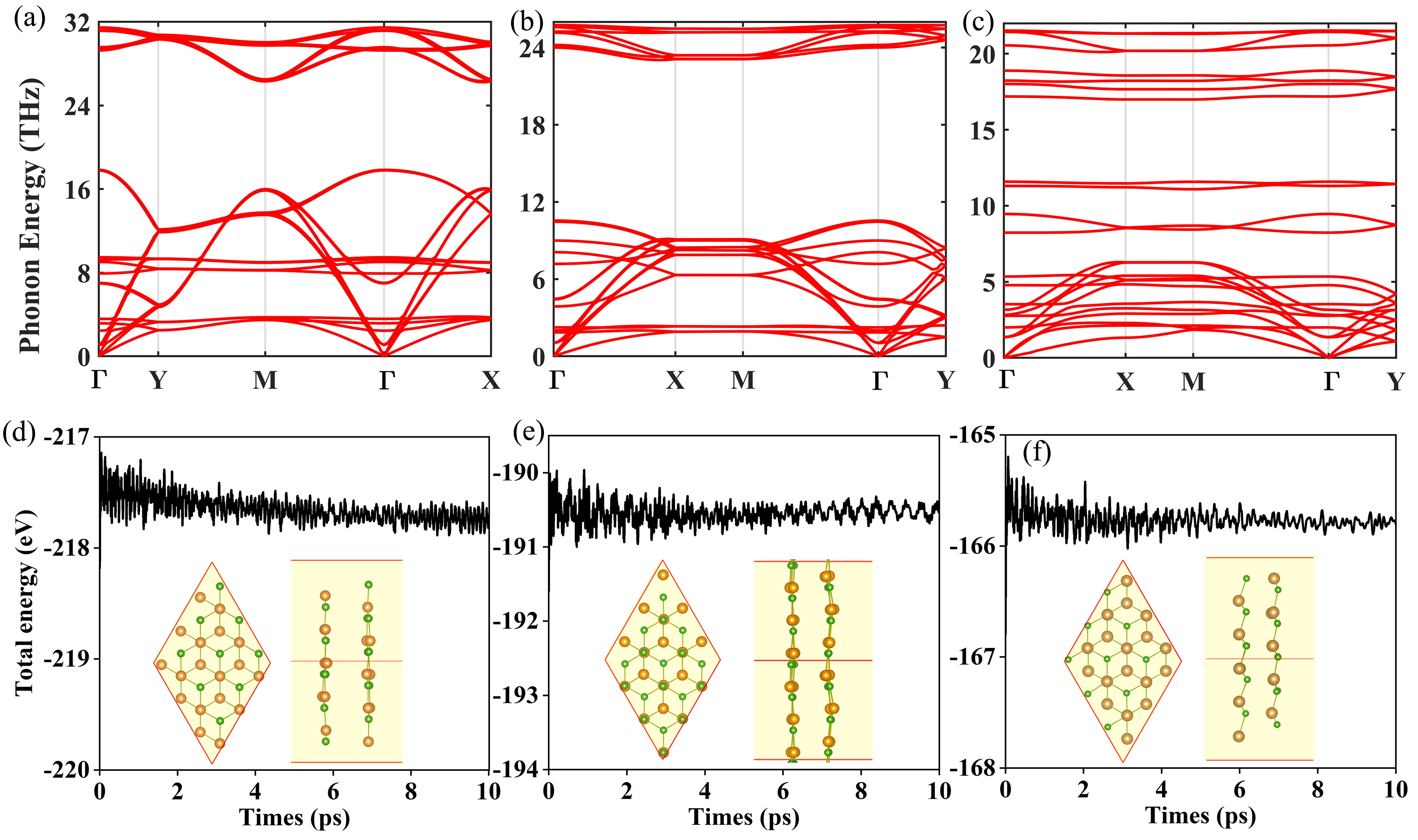}
\centering
\caption{(color online) (a-c) Phonon band dispersion curves of 2D bilayer 3R BP, BAs, and BSb. (d-f) The variation of total energy of each BX bilayer as a function of time when T=500 K. Insets: top and side views of a snapshot for each 3$\times$3 BP supercell at 10 ps.}
\label{fig2}
\end{figure*}

Monolayer BX is a planer hexagonal lattice material similar to graphene, where B atoms and X atoms evenly distribute in the vertices of the hexagon adjacently. There are two typical stacking pattens for bilayer BX, namely 1T stacking and 3R stacking. For the 1T bilayer BX, the top and the bottom layers overlap completely without any dislocation, and thus no vertical electric polarization occurs owning to the mirror symmetry between two layers (see Fig.S1 (a)). While for the 3R stacking, the X atoms in the bottom layer are right below the hexagon centers of the top layer, and the B atoms in the bottom layer are coincidentally below the X atoms in the top layer as shown in Fig.~\ref{fig1}(a). Due to this staggered stacking configuration, the horizontal mirror symmetry between the top layer and the bottom layer is broken, and a spontaneous electric polarization perpendicular to the plane is induced. With increase of interlayer interaction from BP to BSb, the planer bilayer structures gradually evolve into wrinkled ones especially for BSb as shown in Fig.~\ref{fig1}(a), which further increases the spatial reflection asymmetry and induces larger polarization. In order to obtain the polarization direction, the renormalized real space charge density difference $\Delta\rho(\textbf{r})$ was calculated. $\Delta\rho(\textbf{r}) =[\rho_{3R}-\rho_{top}-\rho_{bottom}]/n$, where $\rho_{3R}$, $\rho_{top}$ and $\rho_{bottom}$ are charge density of 3R bilayer BX, individual top and bottom monolayer, respectively, and $n=2$ indicates bilayers. Both the 1D plane integrated $\Delta\rho(z)$ and 3D isosurface plot of $\Delta\rho(\textbf{r})$ are displayed in Fig.~\ref{fig1}(b). Obviously, charge depletes in the top layer, while accumulates in the bottom layer for each 3R BX. Bader charge analysis suggests that there are 0.014, 0.023 and 0.257 electrons transferred from the top layer to the bottom layer for 3R BP, BAs and BSb, respectively. The increased charge accumulation at the interface from BP to BSb indicates more interlayer interaction of the latter than the former. This asymmetry induced interlayer electric polarization has been reported in some other 2D vdWHs and confirmed a 2D out-plane ferroelectric phase.\cite{BN,OOP,VS2} In order to quantitatively characterize the polarization intensity, the magnitude of interfacial polarization density $P$ was calculated based on the Berry phase method, which is roughly equal to the polarized surface charge density multiplied by the interlayer distance.\cite{ACSnano} The calculated values are $P_{BP}=1.6\times10^{-12} C/m$, $P_{BAs}=2.2\times 10^{-12} C/m$ and $P_{BSb}=13.9\times 10^{-12} C/m$, which are comparable or larger than those of bilayer BN  ($2.0\times10^{-12} C/m$)\cite{ZhaoSP} and 1T MoS$_{2}$ ($0.22 \mu C/m^{2}$)\cite{ACSnano11}, indicating strong out-plane ferroelectricity of 3R bilayer BX. The increased $P$ from BP to BSb is attributed to the discontinuity of electrostatic potentials ($\Delta V$) between the top and bottom layers, which is equal to 0.15 V, 0.18 V and 0.47 V, respectively, for 3R BP, BAs and BSb as shown in Fig.S2. The much larger difference of $\Delta V$ between BSb and BAs than that between BAs and BP is attributed to the wrinkled structure of BSb.

In order to confirm the ferroelectric phase of this out-plane polarization, the modest energy barrier $\Phi$ is detected by evaluating the total energies at different transition states for each 3R BX. A series of transition states can be obtained by sliding the top layer along the armchair direction until the bilayer BX returns to its initial 3R configuration. It is easy to find that when the sliding distance is equal to two times of B-X bond length, the final stacking is the same as the top-bottom flipped initial 3R stacking. Using the nudged elastic band (NEB) method, the evolution of total energies along the reaction path is shown in Fig.~\ref{fig1}(c). Due to the interlayer vdW interaction, the reaction energies increase firstly and then decrease symmetrically for all the bilayer BX owning to the same initial and final 3R stacking configurations. In other words, the modest energy barriers of $\Phi_{BP}$=19.5 meV/f.u., $\Phi_{BAs}$=28.4 meV/f.u., and $\Phi_{BSb}$=128.5/f.u. meV should be surmounted during the interlayer translation from the initial 3R stacking to the final 3R stacking. The gradually enlarged energy barriers from BP to BSb can be ascribed to the enhanced interlayer vdW interaction. These values of reaction energy are comparable or lower than that of bilayer 3R MoS$_{2}$ (17.3 meV/f.u.),\cite{scireport} $\alpha-$In$_{2}$Se$_{2}$ (50 meV/f.u.),\cite{OOP40} functionalized bismuth layer (120-548 meV/f.u.),\cite{OOP41} and $\gamma-$GeSe (887 meV/f.u.),\cite{OOP42}), implying entirely feasible experimental feasibility.

\begin{figure*}[!tb]
\includegraphics[width=15cm]{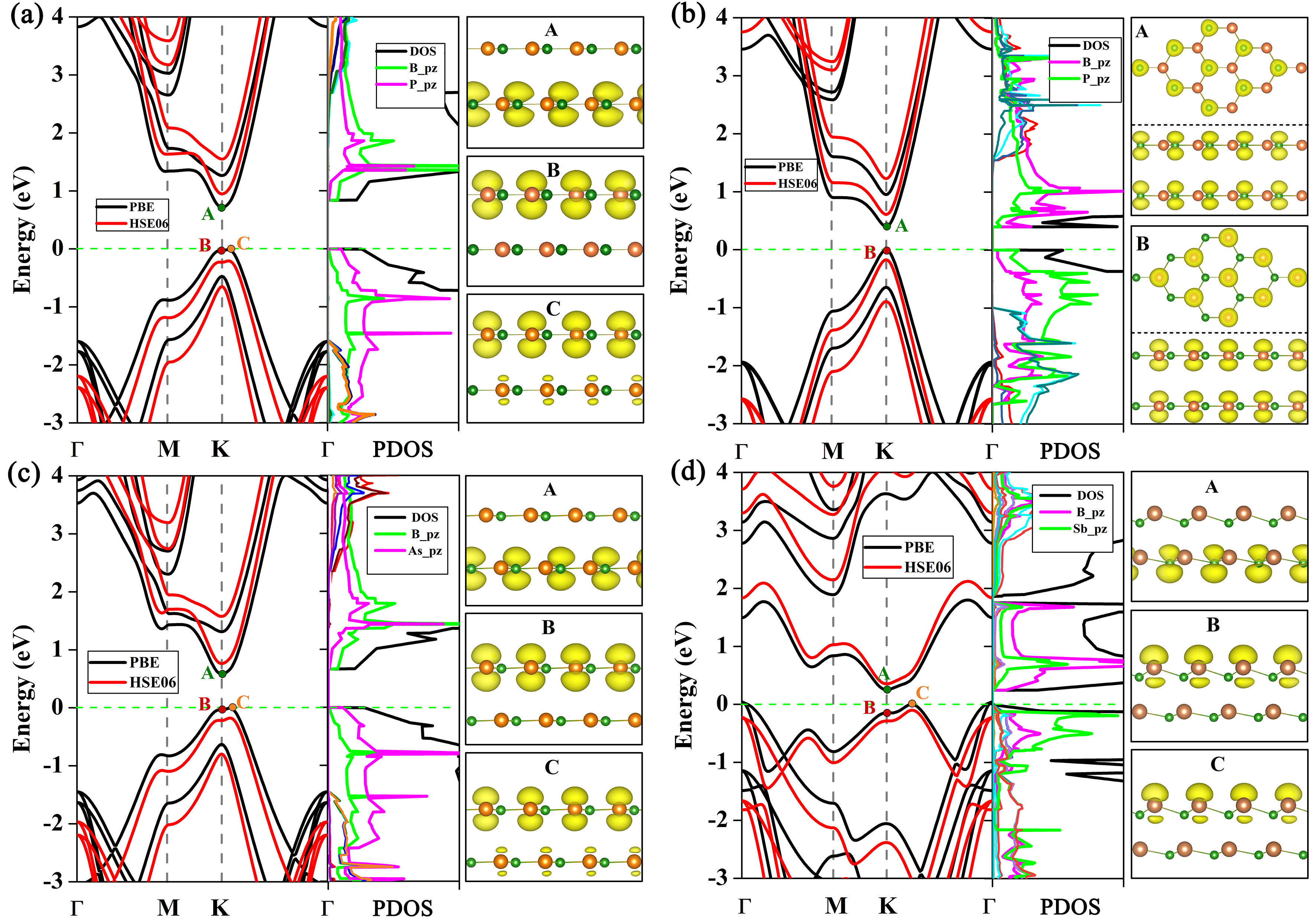}
\centering
\caption{(color online) Band structures (left panels), PDOS (middle panels), and partial charge densities at the CBM and VBM (right panels) of 3R bilayer BP (a), 1T bilayer BP (b), 3R bilayer BAs (c), and 3R bilayer BSb (d), respectively. In each panel, the Fermi level of each system is set to zero and marked by the green vertical dotted lines. The band structures at PBE (HSE06) levels are labeled by black (red) lines and the value of the isosurface in each middle panel is set at 0.0007 e{\AA}$^{-3}$.}
\label{fig3}
\end{figure*}

Furthermore, the ferroelectric switching of the 3R BX structures along the reaction path is shown in Fig.~\ref{fig1}(d). Owing to the top-bottom reversion between the initial and the final stackings, the magnitudes of interfacial dipole moments are centrosymmetric along the total reaction path indicating a reversion of out-plane ferroelectricity. Based on the obtained $P$, the interlayer voltage $U$ can be estimated coarsely by using the formula $U=P/\varepsilon_0$, where $\varepsilon_0$ is vacuum permittivity. For 3R stacking BX, $U$ is equal to $\pm 0.12 V$, $\pm 0.15 V$ and $\pm 0.57 V$, respectively, for BP, BAs and BSb. By dragging either monolayer along the armchair direction, these interlayer voltages can be reversed periodically. Such intriguing vertical ferroelectric switching properties can be used as a nanogenerator, where the electron flow and energy harvesting will be manipulated with the mechanism akin to the previously designed triboelectric nanogenerator.\cite{ACSnano21,ACSnano22,ACSnano23}

%\begin{figure*}[!tb]
%\includegraphics[width=16cm]{FigS1.jpg}
%\centering
%\caption{ (color online) Charge density deformation of plane projection and three-dimensional integration for the 2D bilayer 3R (a) BP, (b) BAs, and (c) BSb bilayers, respectively. In each panel, the red horizontal dotted lines denoted the Fermi level, the green vertical lines indicate edges of the top and bottom layers, and the yellow and cyan isosurfaces represent the charge accumulation and depletion at the interface with an isovalue of 0.015 e{\AA}$^{-3}$ (a, b) and 0.1 e{\AA}$^{-3}$ (c), respectively.}
%\label{figS1}
%\end{figure*}

%\begin{figure*}[!tb]
%\includegraphics[width=16cm]{FigS4.jpg}
%\centering
%\caption{ (color online) Plane averaged electrostatic potential of bilayer 2D 3R (a) BP, (b) BAs, and (c) BSb along the z direction. In each panel, the potential differences between top and bottom layers of the corresponding crystals are marked by $\Delta V$.}
%\label{figS4}
%\end{figure*}

\subsection{Crystal stabilities and electronic properties of 3R BX (X=P, As, Sb)}

Before further evaluating the electric properties of ferroelectric 3R bilayer BX, their structural stability should be examined firstly. Firstly, the real space projection of electron localized functions (ELFs) is calculated to characterize the bonding types of 3R BX. The magnitude of ELF from 0 to 1 describes the spatial localization extent of the reference electrons, where ELF = 1 or 0 represents complete localization or full decentralization of the electrons. For each 3R bilayer BX, substantial concentration of electrons is located between B and X atoms as shown in Fig.S3, indicating strong covalent bonding feature and stable configuration. Next, the interfacial binding energy is calculated to examine the energy stability of the 3R bilayer BX by $E_{b}=(E_{3R}-E_{top}-E_{bottom})/S$, where $E_{3R}$, $E_{top}$ and $E_{bottom}$ are energies of the unit cell of the 3R bilayer BX, the top and the bottom monolayer, respectively, and $S$ is the area of unit cell. The interfacial binding energies are found to be $E_{b, BP}=-0.27 J/m^{2}$, $E_{b, BAs}=-0.35 J/m^{2}$ and $E_{b, BSb}=-0.12 J/m^{2}$, which are comparable to those of some recently reported vdWHs including WTe$_{2}$/HfS$_{2}$ ($-0.204 J/m^{2}$)\cite{jpcc121}, phosphorene/SnSe$_{2}$ ($-0.019 J/m^{2}$),\cite{PRA}, KAgSe/SiC$_{2}$ ($-0.327 J/m^{2}$),\cite{WQJMCC} \emph{et.al.}. Moreover, the stability of 3R bilayer BX is further confirmed by phonon spectrum and \emph{ab initio} molecular dynamics simulations. As shown in Fig.~\ref{fig2} (a-c), no imaginary phonon mode is observed in the entire Brillouin zone, confirming the dynamical stability of 3R BX structures. Finally, the temporal evolution of the total energies of 3$\times$3 3R BX supercells are evaluated to examine their thermal stabilities. The temperature is set equal to 500K and the evolutionary time is up to 10 ps with a time step of 1 fs. As shown in Fig.~\ref{fig2} (d-f), the total energy of each 3R BX oscillates around a fixed value with only small fluctuations, indicating the thermal stability of each 3R BX during the temporal evolution. Meanwhile, the structural integrity of dynamic configurations maintains well without broken bond or geometry reconstruction, revealing excellent thermal stability of each 3R BX structure even at high temperature. In summary, the structural stability of the three 3R bilayer BX is confirmed from several aspects, which provides a possible guidance for their experimental synthesis.

%\begin{figure*}[!tb]
%\includegraphics[width=14cm]{FigS3.jpg}
%\centering
%\caption{Top and side views of the ELF distributions along arbitrary in-plane directions of 2D bilayer 3R (a) BP, (b) BAs, and (c) BSb, respectively.}
%\label{figS3}
%\end{figure*}
 %to gain more insight into the different electronic behaviors between bilayer 3R and 1T BP structures

\begin{figure*}[!tb]
\includegraphics[width=15cm]{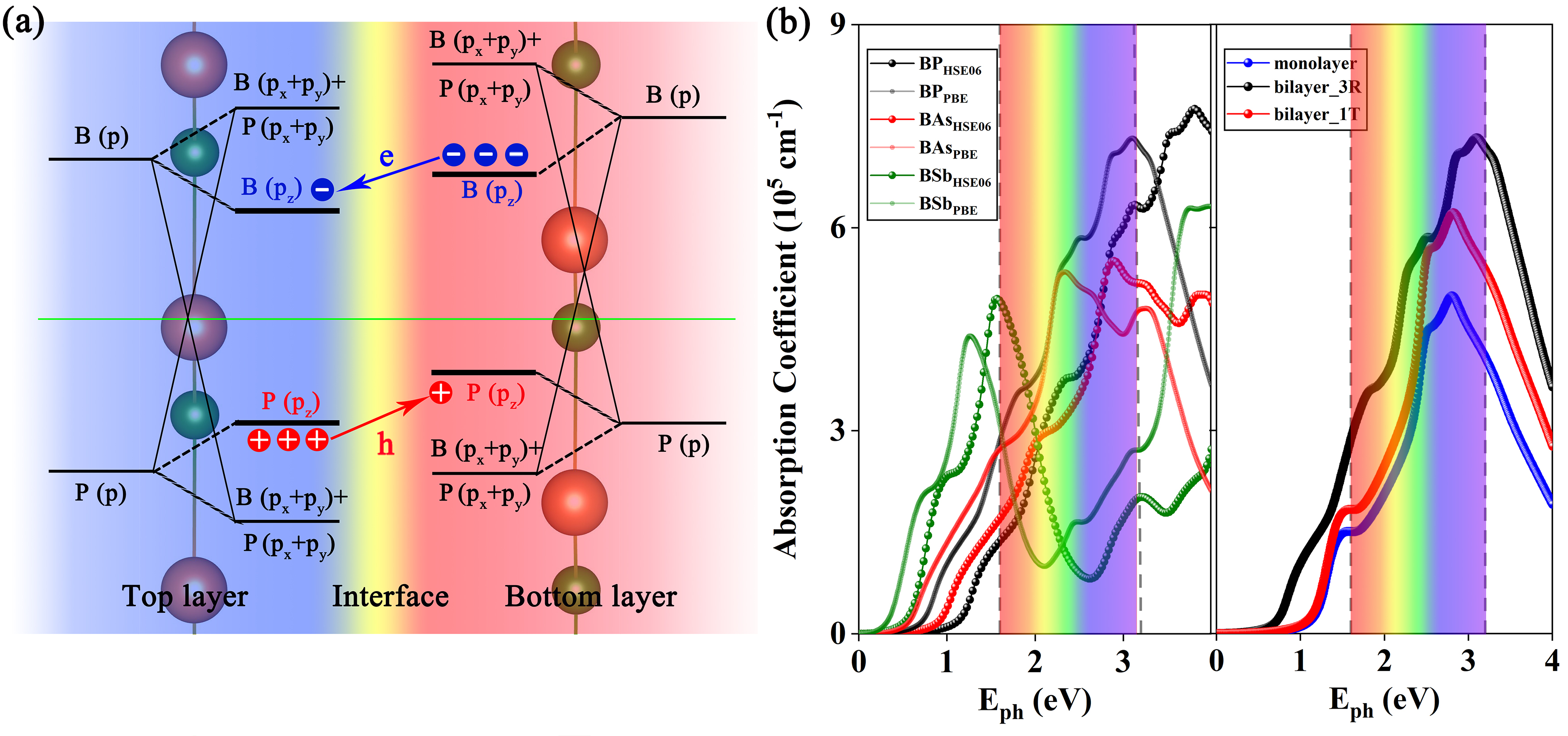}
\centering
\caption{(color online) (a) Schematic diagrams of the energy levels at the $K$ point of 2D bilayer 3R BP, where the Fermi level is marked by green horizontal line. (b) Optical absorption coefficients versus energy of the incident light for different 2D bilayer 3R BX (X=P, As, Sb) systems (left panel) and different configurations of BP (right panel). Here, the polarization vector is set parallel to the plane and along the zigzag edges. The region of visible light in each panel are denoted by the colored translucent shadow areas between two vertical dash lines.}
\label{fig4}
\end{figure*}

In the following, we examine the electronic behaviors of the 3R bilayer BX. Fig.~\ref{fig3} (a) shows the band structure, projected density of states (PDOS) and partial charge densities of 3R bilayer BP with the exchange-correlation interaction considered at both PBE and HSE06 levels. Obviously, the band structures based on both exchange-correlation interactions are qualitatively consistent. The 3R bilayer BP shows indirect band gap semiconducting behavior with band gap equal to 0.73 eV at PBE level and 1.15 eV at HSE06 level. The moderate band gap at HSE06 level is very close to the optimum range of 1.2$\sim$1.6 eV for excitonic solar cells (XSCs).\cite{WQKAS18} The conduction band minimums (CBM) is located at K point, while the valence band maximum (VBM) is slightly deviated away from K point accompany with tiny upward from B point to C point. Considering that the bands of bilayer should be roughly a superposition of those from two individual monolayers (see Fig.S1(b)) due to very weak vdWH interaction, we expect that this indirect band gap can be ascribed to the interlayer built-in electric field caused by the spontaneous ferroelectric polarization. To confirm this conjecture, the electronic behaviors of the 1T bilayer BP is also plotted in Fig.~\ref{fig3} (b). Obviously, direct band gap behavior is observed for either PBE or HSE calculation, where both CBM and VBM are located at K point. Its band gap is equal to 0.42 eV at PBE level and 0.78 eV at HSE06 level, which is smaller than that of corresponding 3R bilayer BP, indicating the stronger interlayer vdWH interaction of 1T BP. As shown by PDOS in the middle panels of Fig.~\ref{fig3} (a) and (b), the CBMs are dominated by the B-$p_{z}$ orbitals, while the VBMs are mainly contributed by the P-$p_{z}$ for both 3R and 1T bilayer BP indicating their similarity. However, further calculation of partial charge densities as shown in the right panels of Fig.~\ref{fig3} (a) and (b) points out their difference. For the 1T bilayer BP, both the CBM and the VBM are equally contributed by the top layer and the bottom layer, and therefore no interlayer charge polarization. While, for the 3R bilayer BP, the CBM is mostly dominated by the bottom layer, and the VBM is mainly contributed by the top layer, indicating an obvious spatial separation between the CBM and the VBM. As a result, a type-II band alignment is formed for the 3R bilayer BP owing to the real space charge separation between CBM and VBM. Similar behavior of type-II band alignment and real space electron-hole separation are also found for 3R bilayer BAs and BSb as shown in Fig.~\ref{fig3} (c) and (d), respectively, apart from the enlarged delocalizations of VBM and the decreased band gaps due to the stronger electric polarization. In Fig.~\ref{fig4} (a), we provide a more essential and intuitive physical image to present the type-II band alignment of 3R BP, where the CBM and VBM of top and bottom layer are dominated by B-$p_{z}$ and P-$p_{z}$, respectively. Obviously, driven by the interlayer built-in electric field, the energy levels of B-$p_{z}$ and P-$p_{z}$ from the top layer are higher than those of the bottom layer.

For this type-II 3R bilayer BP ferroelectric system, a series of intriguing photovoltaic excellences will be formed naturally. (1) Different from 1T bilayer BP, the built-in electric field in 3R bilayer BP will facilitates the separation of the free electrons and holes into different layers, accompanied with reduction of the photo-excited electron-hole recombination opportunity.  (2) As for the heterojunctions with distinct 2D materials, lattice mismatch would induce large potential barriers which limits their stability and experimental synthesis. In addition, the possible type-I or type-III band alignments between layers may restrict their application in photovoltaic field someway.\cite{WQJMCA, WQJMCC} Instead, the 3R bilayer BP possesses identical component materials, intrinsic type-II band alignment and nearly complete real space electron-hole separation. (3) The unique vertical ferroelectric switching properties enable the realization of electron-hole separation and reversion in real space by relative translation between layers. These properties of 3R bilayer BP are appealing for experimental design, application and manipulation in the field of optoelectronic devices and solar energy conversions.\cite{cheng20182d} More strikingly, these ideal performances of ferroelectric photovoltaic materials can also be extended to other 2D layered crystals beyond the 3R bilayer BX, hence providing a new strategy for advancing the fundamental research, experimental design and unique industrial applications of future photovoltaic conversion devices.

\begin{figure*}[tb]
\includegraphics[width=15cm]{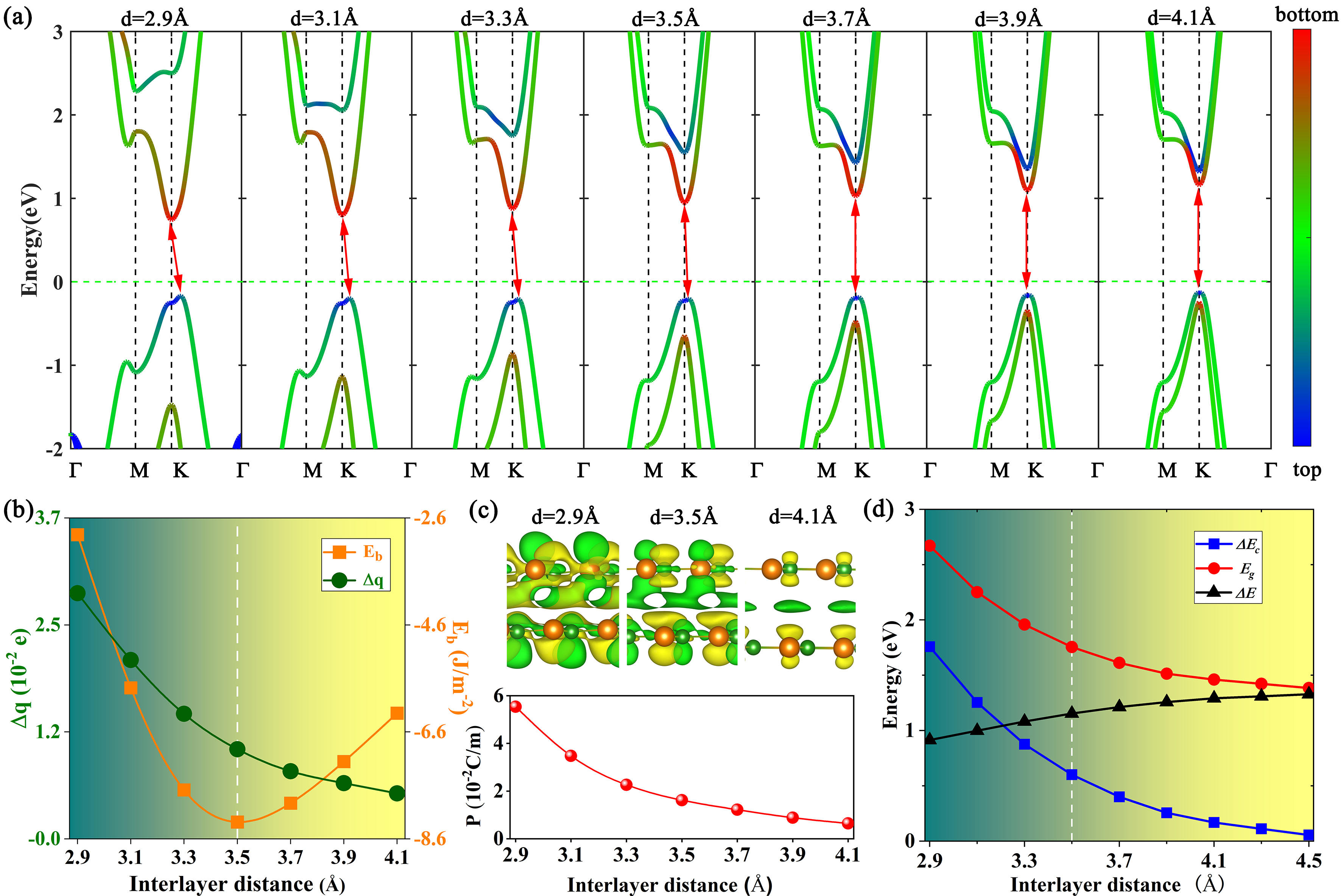}
\centering
\caption{Band structures of the 3R bilayer BP under various interlayer distances $d$ at HSE06 level. In each panel, the bands are projected to the top and the bottom layers with different magnitudes indicated by different colors. The green horizontal lines denote the Fermi level, and the red arrows indicate the indirect-to-direct band gap variation. (b) Electron transfer ($\Delta$q) from the top layer to the bottom layer and binding energy (E$_{b}$) as functions of $d$. (c) Top panel: isosurface of charge density difference under three different $d$. The yellow and green regions correspond to charge accumulation and depletion, respectively. Bottom panel: dipole moment $P$ changes over $d$. (d) Band gap $\Delta E$, conduction band offset $\Delta E_c$ and band gap of the donor segment $E_g$ versus interlayer distance $d$. In (b) and (d), The white vertical line represents the equilibrium interfacial distance of the system..}
\label{fig5}
\end{figure*}

\subsection{Carrier mobilities and optical absorptions of 3R BX (X=P, As, Sb)}

In light of potential application as 2D photoluminescence device, the carrier mobility and optical absorption coefficient are two fundamental factors to be explored. The carrier mobilities of the 3R bilayer BX are calculated based on the deformation potential (DP) theory, which can be expressed as follows,\cite{JMCA631}
\begin{equation}
\mu_{2D}=\frac{e\hbar^{3}C_{2D}}{k_{B}T|m_{e/h}^{*}|(E_{1})^{2}}.
\label{cur1}
\end{equation}
where $C_{2D}$ is the elastic modulus, $k_{B}$ is the Boltzmann constant, $T$ is the temperature, $m_{e/h}^{*}$ is the effective mass of electrons or holes, and $E_{1}$ is the deformation potential. This model of deformation potential theory has been used extensively to examine the $\mu_{2D}$ of 2D atomic structures. As shown in Table~\ref{tab:table1}, ultra-high carrier mobilities have been found for the proposed 3R bilayer BX structures. For example, the maximum $\mu_{2D}$ of the 3R bilayer BP can reach $5.72\times 10^{5} cm^{2}\cdot V^{-1}\cdot s^{-1}$ for electrons along armchair direction. These ultra-high carrier mobilities of 3R bilayer BX can further highlight their potential application prospects in electronics and photoelectronics.

\begin{table}[ht]
  \centering
  \caption{ carrier mobilities ($\mu _{2D}$) of bilayer 3R BX crystals for electrons and holes along armchair and zigzag directions.}
    \begin{tabular}{lccccccccc}
    \hline
    \multirow{3}{*}{crystal structures}  &  \multicolumn{2}{c}{zigzag} & & \multicolumn{2}{c}{armchair}   \\
    \cline{2-3}
    \cline{5-6}
    & \multicolumn{1}{c}{e} & \multicolumn{1}{c}{h} & & \multicolumn{1}{c}{e} & \multicolumn{1}{c}{h}\\
    \hline
    $BP  (10^{4} cm^{2}V^{-1}s^{-1})$  &  5.72   & 1.26   & &  3.73  &  1.46  \\
    $BAs (10^{4} cm^{2}V^{-1}s^{-1})$  &  2.49   & 1.56   & &  1.07  &  1.85  \\
    $BSb (10^{4} cm^{2}V^{-1}s^{-1})$  &  1.44   & 1.41   & &  0.62  &  1.61  \\
    \hline
    \end{tabular}%
  \label{tab:table1}%
\end{table}%

\begin{figure*}[tb]
\includegraphics[width=15cm]{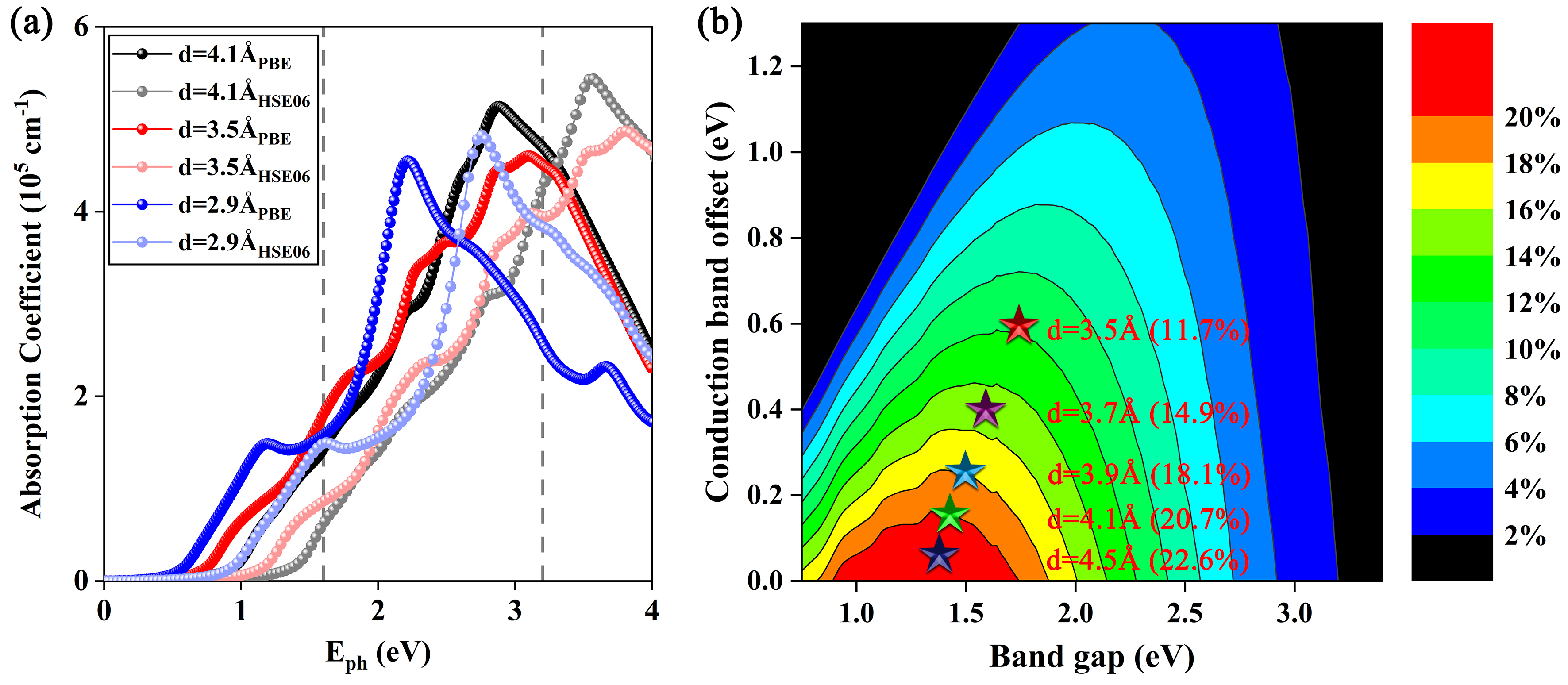}
\centering
\caption{(color online) (a) Optical absorption coefficients $a(\omega)$ versus energy of incident light for 3R bilayer BP under different interlayer distances $d$. The polarization vector is set parallel to the plane and along the zigzag edges. The two vertical dash lines indicate the visible light region. (b) Phase diagram of PEC as functions of donor bandgap $E_g$ and conduction band offset $\Delta E_c$ under different interlayer distance $d$.}
\label{fig6}
\end{figure*}

The optical absorption coefficient $a(\omega)$ is calculated based on the following formula,\cite{JMCA560}
\begin{equation}
a(\omega) = \frac{\sqrt{2\omega}}{c} \sqrt{\sqrt{\epsilon_1^2(\omega)+\epsilon_2^2(\omega)}-\epsilon_1(\omega)},
\end{equation}
where $\omega$ is the frequency of incident light, $c$ is the light velocity, $\epsilon_1$ and $\epsilon_1$ are the real and imaginary parts of frequency-dependent dielectric function, respectively. $\epsilon_1$ can be calculated by Kramers-Kronig transformation, and $\epsilon_2$ is the summation over the empty states.\cite{JMCA560} In this calculation, the directions of polarization vectors of the incident light are set parallel to the plane. As shown in the left panel of Fig.~\ref{fig4} (b), $a(\omega)$ maintains zero for each 3R BX until $\omega$ of the incident light is larger than the cut-off frequency, and then increases versus the energy of incident light. Besides, the curves of $a(\omega)$ obtained at HSE06 level are qualitatively consistent with the curves obtained at PBE level with proper blue shifts, which is accordant with the information given by the band structures as shown in Fig.~\ref{fig3}. Intriguingly, all $a(\omega)$ curves possess wide peaks over the range of visible light ($\sim1.6-3.2 eV$). The maximum values can reach $\sim4.95-7.75 \times10^{5} cm^{-1}$ for 3R BP, BAs, and BSb, respectively, which are one order of magnitude larger than those of ideal 2D layered KAgSe and TiNX (X=F, Cl, Br).\cite{WQKAS1, WQKAS18} To further understand the mechanism of the large absorption coefficients of these 3R bilayer BX, the absorption spectrums of the monolayer and 1T bilayer BP are also evaluated to make a comparison as shown in the right panel of Fig.~\ref{fig4} (b). Obviously, $a(\omega)$ of the 3R bilayer BP is the largest over the visible light range, indicating most excellent optical absorption performances.

\subsection{Tunable effects of interfacial distance on the 3R bilayer BP }

In the following, we regulate the electric and photoelectric properties of the 3R bilayer BX by changing their interfacial distance, which can be implemented through applying a vertical strain by vacuum thermal annealing or imposing pressure with a scanning tunneling microscopy tip in experiments.\cite{JMCC64, JMCC65} As shown in Fig.~\ref{fig5} (a), when the interfacial distance $d$ of the 3R bilayer BP increases from 2.9 {\AA} to 4.1 {\AA}, the band gap increases from 0.91 eV to 1.29 eV (HSE06), accompany with a conversion from indirect band gap to direct band gap. This variation of band gap is desired to enhance the photovoltaic performance of the material. To understand the variation of band gap, charge transfer $\Delta q$ and binding energy $E_{b}$ under different $d$ are evaluated and presented in Fig.~\ref{fig5} (b). $E_{b}$ remains negative within the entire range of $d$, indicating the energetic stability of the 3R bilayer BP under all vertical strain, and the minimum value appears at the equilibrium distance $d=3.5$ {\AA}. Meanwhile, $\Delta q$ decreases monotonically versus $d$, revealing declined charge transfer from the top layer to the bottom layer. More visually, real space charge transfer under three typical interfacial distances are shown in the top panel of Fig.~\ref{fig5} (c), where charge accumulation at the interface attenuates with increase of interlayer distance, indicating decaying interfacial interaction. It is the decaying interfacial interaction that induces monotonically decreased dipole moment $P$ as shown in the bottom panel of Fig.~\ref{fig5} (c) and therefore the variation from indirect band gap to direct band gap.

Next, we explore the tunable photovoltaic performances of the 3R bilayer BP under various vertical strain. Fig.~\ref{fig6} (a) presents the optical absorption coefficients $a (\omega)$ under different interlayer distance at both PBE and HSE06 levels. The curves obtained at HSE06 level are roughly accordant with the blue-shifted curves obtained at PBE level with qualitative agreement. More intriguingly, both magnitudes and positions of the peaks of $a (\omega)$ are changed obviously under different vertical strains. When $d$ is enlarged to 4.1 {\AA}, larger magnitude of peak can be observed over the visible light range, revealing the stronger capture performance of visible light. In contrast, when $d$ is compressed to 2.9 {\AA}, a significant red-shift is generated due to the band recombination under vertical strain. PCE is another important factor to describe the efficiency of XSC, which can be calculated by \cite{JMCA60},
\begin{equation}
PCE = \frac{0.65(E_{g}-\Delta E_{c}-0.3)\int_{E_{g}}^{\infty}\frac{P(\hbar \omega)}{\hbar \omega}d(\hbar \omega)}{\int_{0}^{\infty}P(\hbar \omega)d(\hbar \omega)},
\end{equation}
where 0.65 is the assumed value of the band-fill factor; $E_g$ is the band gap of the donor segment; $\Delta E_c$ is the magnitude of conduction band offset; $(E_{g}-\Delta E_{c}-0.3)$ is the estimated value of the maximum open circuit voltage; and $P(\hbar \omega)$ is the $AM1.5$ solar energy flux with the photon energy of $\hbar \omega$. As shown in Fig.~\ref{fig5} (d), the band gap increases gradually with stretch of interlayer distance $d$, while $E_g$ and $\Delta E_c$ decrease monotonically due to the reduced vdWH interaction. As a result, PCE increases with interlayer distance. Fig.~\ref{fig6} (b) shows the phase diagram of PCE versus $E_g$ and $\Delta E_c$ of the 3R bilayer BP under different interlayer distance. At the equilibrium interlayer distance (3.5 {\AA}), $E_g$ and $\Delta E_c$ are 1.74 eV and 0.60 eV, respectively, and PCE can reach 11.7\%. This large PCE is coincidentally equal to the best certified efficiency of the organic solar cell (11.7\%)\cite{JMCA61}, and is larger than that of the typical MoS$_{2}$/p-Si heterojunction solar cell (5.23\%)\cite{JMCA62}. When $d$ is widened from 3.5 {\AA} to 4.5 {\AA}, PCE grows from 11.7\% to 22.6\%. Such high PCE has outpaced those of most available materials so far\cite{zhoubinJMCC}, such as recently reported TiNX$_{1}$/TiNX$_{2}$ XSCs (X$_{1}\neq X_{2}$=F, Cl, Br ) ($\sim18\%-22\%$).\cite{WQKAS18} These excellent tunable performances of solar absorption and PCE indicate that our new designed 3R bilayer BP structure can behave as ideal XSCs.

\begin{figure*}[tb]
\includegraphics[width=15cm]{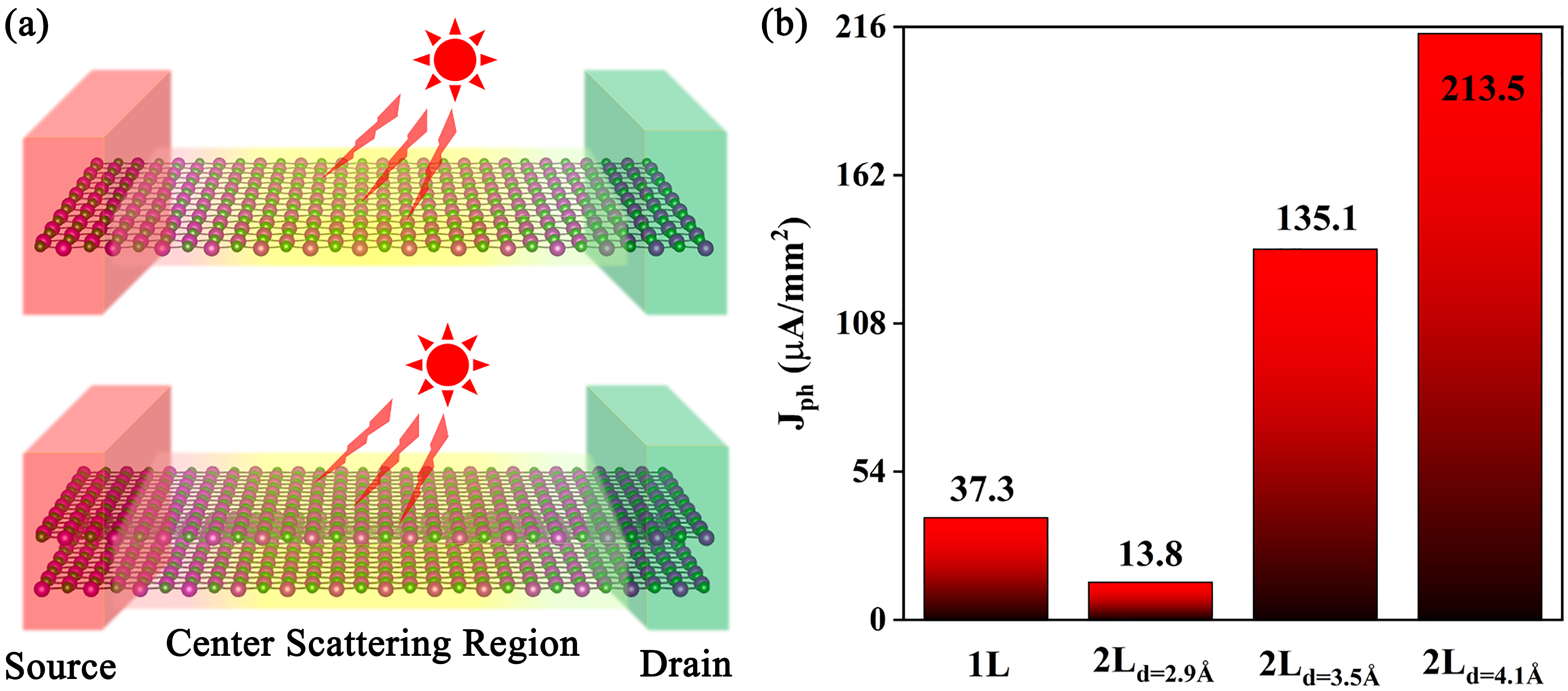}
\centering
\caption{(color online) (a) Schematic structures of 2D two-probe devices based on the monolayer BP (top panel) and the bilayer 3R BP (bottom layer). The drain and source are marked by the left red box and right green region, respectively. The yellow zone in the center scattering region stands for lighting area. (b) Photocurrent of the two-probe systems based on the monolayer (1L) and bilayer BP under different vertical strains. The incident photon energy of each system is set equal to the corresponding band gap.}
\label{fig7}
\end{figure*}

\subsection{Photocurrent in different types of 2D BP based nanodevices}

Although favorable photovoltaic properties of 2D 3R bilayer BX has been discovered, however, it is still necessary to explore the practical photovoltaic performances in 3R bilayer BP based electronic devices considering of the quantum scattering in nanoscale devices due to aperiodic characteristics. Fig.~\ref{fig7} (a) shows two schematic structures of pure monolayer and 3R bilayer BP based 2D two probe devices, where both leads are constructed by periodic extension of the scattering region. Compared with the traditional metal-simiconductor tunneling junctions, there are more advantages of these pure BP devices including simple structure, lattice matching, smooth and continuous interface and easily to be prepared. As reported in previous works, such simplified device model has been confirmed effective to evaluate the photo-electronic transport in practical devices,\cite{WQJMCA,WQKAS1,WQKAS18} and has been employed extensively in experimental battery design\cite{WQJMCC,PRA,WQJMCA,WQKAS1,WQKAS18}. For each nanodevice shown in Fig.~\ref{fig7} (a), the whole scattering region is irradiated by a perpendicular linearly polarized light, and a small bias voltage is applied between the source and drain to drive the flowing of photocurrent. When the photon energy is greater than the band gap, electrons are excited from the valence band to the conduction band accompany with the generation of electron-hole pairs. Under the driven force of potential difference between source and drain, the electron-hole pairs will be further separated to electrons and holes transferring along opposite directions, and finally a photocurrent occurs. In our calculation, the power density of incident light is set to 10$^{3}$ $\mu W/mm^{2}$, and the bias voltage is equal to 0.2 V, which is far less than the band gap of BP to ensure only the photocurrent is generated. At the first order of Born approximation, the photocurrent flowing into the left probe can be expressed as follows,\cite{JMCA65,JMCA66,JMCA67}
\begin{equation}
\begin{aligned}
I_{L}^{ph}=\frac{ie}{h}\int \textbf{Tr}\left[\Gamma_{L}\{G^{<(ph)}+f_{L}(E)(G^{>(ph)}-G^{<(ph)})\}\right]dE,
%J_{L}^{ph}=&\frac{ie}{h}\int \textbf{Tr}\left[\Gamma_{L}\{G^{<(ph)}+f_{L}(E)(G^{>(ph)}-G^{<(ph)})\}\right]dE,
\end{aligned}  \label{cur0}
\end{equation}
where $\Gamma_L$ is the line-width function, $f_L$ is the Fermi distribution function, and $G^{</>(ph)}$ is the lesser/greater Green's function of the two probe system including electron-photon interaction.\cite{JMCA67} More detailed theoretical formalism of photocurrent can be found in Ref.[\citenum{JMCA66}].

Fig.~\ref{fig7} (b) shows the photocurrent density $J_{ph}=I_L^{ph}/S$ of monolayer BP and 3R bilayer BP nanodevices with three different interlayer distances $d$ when the incident photon energy $E_{ph}$ is equal to the band gap $\Delta E$ of corresponding BP. $J_{ph}$ of different BP devices can reach 13.8-213.5 $\mu A/mm^{2}$, which are large enough for experimental detection. For the 3R bilayer BP device at equilibrium interlayer distance (3.5 {\AA}), $J_{ph}$ is roughly 3.5 times larger than that of the monolayer BP device due to the stronger absorption ability and type-II band alignment. For the bilayer 3R BP devices, $J_{ph}$ is largely improved from 13.8 $\mu A/mm^{2}$ to 213.5 $\mu A/mm^{2}$ with increasing $d$ from 2.9 {\AA} to 4.1 {\AA}. This phenomenon can be attributed to increasing PCE and conversion from indirect to direct band gap as functions of $d$, considering of their similar optical absorptions and type-II band alignments. Compared to the indirect band gap systems, the excitation of electrons in the direct band gap devices no longer needs the assistance of phonons, hence suggesting a much larger photocurrent. On the contrary, the obvious indirect band gap and ultra-low PCE of the bilayer device with $d$=2.9 {\AA} reduce $J_{ph}$ even smaller than that of the monolayer device. Note that all results in Fig.~\ref{fig7} (b) are calculated at PBE level considering of the huge computational cost of photocurrent in HSE06 calculation. We believe that the photocurrents obtained at HSE06 level should be accordant qualitatively with those obtained at PBE level with quantitative difference.

\begin{figure*}[tb]
\includegraphics[width=15cm]{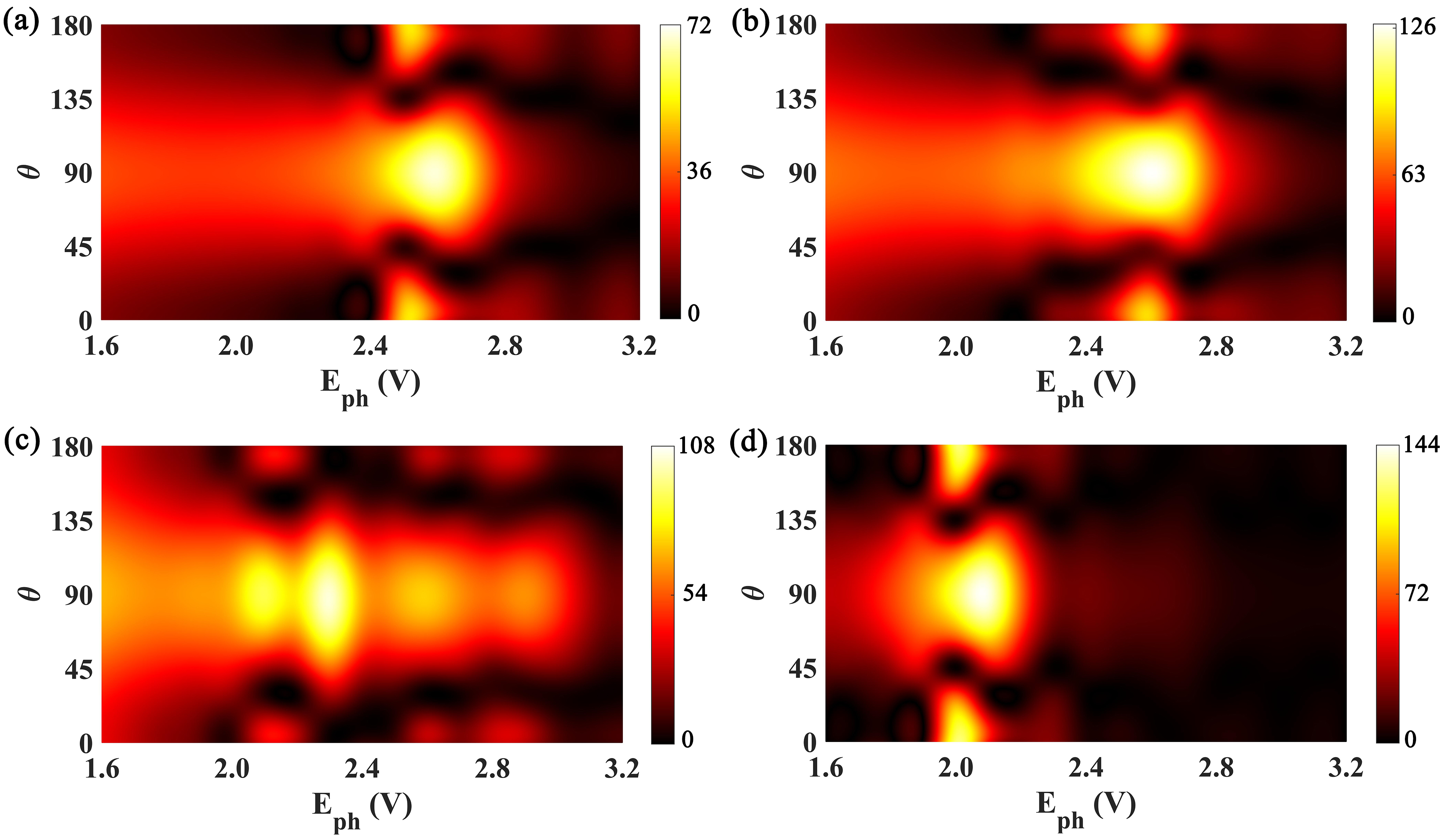}
\centering
\caption{(color online) Photocurrent density J$_{ph}$ versus incident light energy E$_{ph}$ and polarization angle $\theta$, for (a) monolayer BP and bilayer 3R BP with interlayer distance $d$ is equal to (b) 4.1 {\AA}, (c) 3.5 {\AA}, and (d) 2.9 {\AA} at PBE level. The power density P$_{in}$ of the incident light is 10$^{3}$ $\mu Wmm^{-2}$.}
\label{fig8}
\end{figure*}

In the following, we discuss the variation of $J_{ph}$ \emph{versus} energy $E_{ph}$ and polarization angle $\theta$ of linearly polarized incident light. As shown in Fig.~\ref{fig8}, broad peaks of $J_{ph}$ are detected in the visible light region with the maximum value equal to of 72 $\mu A/mm^{2}$ for monolayer BP device and 108-144 $\mu A/mm^{2}$ for three 3R bilayer devices. These energies of maximum $J_{ph}$ roughly correspond to the peaks of optical absorption coefficient a($\omega$) as shown in Fig.~\ref{fig6} (a), and are also close to the energy difference between the two maximum values of the density of states near the Fermi energy (see Fig.S4). The detected maximum values of $J_{ph}$ are far larger than those of the MoS$_{2}$ and KAgSe devices under the same external conditions\cite{JMCA68,WQKAS1}, indicating an appealing photoelectric conversion performances of these 2D BP devices in the visible light range. For three 3R bilayer BP devices, an obvious red-shift of $J_{ph}$ peak is observed when interlayer distance is stressed from 4.1 {\AA} to 2.9 {\AA}.  In addition, $J_{ph}$ of each BP device is roughly symmetrical relative to $\theta=90^{\circ}$ with the maximum values locating at $\theta=90^{\circ}$. This is reasonable because $J_{ph}$ can be expressed as a linear superposition of three terms, which are proportional to $\sin^{2}\theta$, $\cos^{2}\theta$, and $\sin2\theta$, respectively, and the final magnitude of $J_{ph}$ is determined by the competition of these three trigonometric terms.\cite{JMCA66} In view of the enlarged band gaps and right-shift a($\omega$) peak under HSE06 calculations, we predict that the peaks of $J_{ph}$ obtained at HSE06 level should be pushed toward the ultraviolet direction as what happened in the previous reports.\cite{WQKAS1}

%\begin{figure*}[!tb]
%\includegraphics[width=16cm]{FigS5.jpg}
%\centering
%\caption{(color online) The density of states (Dos) versus energy of the monolayer BP (red curve) and bilayer 3R BP (green curves) structures under d=4.1{\AA}, 3.5{\AA}, and 2.9{\AA}, respectively. The green dash lines in each panel indicate the fermi levels.}
%\label{figS5}
%\end{figure*}

Finally, the photon responsivity $R_{ph}$ and external quantum efficiency $\tau_{eqe}$ are evaluated based on $J_{ph}$, which are also essential parameters in describing the photovoltaic performance in experiments. $R_{ph}$ is defined by the ratio of  $J_{ph}$ to the incident photon flux $F_{ph}$,\cite{xxx11,xxx22}
\begin{equation}
R_{ph}=\frac{J_{ph}}{eF_{ph}},
\end{equation}
where $e$ is the electron charge, and the photon flux $F_{ph}=P_{in}/(e\cdot E_{ph})$ describes the number of incident photons pear unit area and time. $\tau_{eqe}$ describes the ratio of the excited photoelectrons to the incident photons, which can be calculated by\cite{xxx22,WQJMCA}
\begin{equation}
\tau_{eqe}=R_{ph}E_{ph}/e.
\end{equation}
%and $\eta$ is defined by the ratio of maximum output photoelectron power $P_{out}$ density to the incident one $P_{in}$:
%\begin{equation}
%\eta=P_{out}^{max}/P_{in}
%\end{equation}
%Here, $P_{out}$ of the devices can be obtained by the product of $V_{d}$ and the corresponding $J_{ph}$.
In our calculation, the maximum values of aforementioned parameters are located at $E_{ph}$ = 2.62 eV, 2.61 eV, 2.31 eV, and 2.11 eV for monolayer and 3R bilayer BP with d=4.1{\AA}, 3.5{\AA}, and 2.9{\AA}, respectively. In Table~\ref{tab:table2}, the $R_{ph}$ and $\tau_{eqe}$ are listed including our designed BP devices and previously reported experimental and theoretical parameters. We can find that these parameters in our designed BP devices are comparable and even larger than most of the previous reported ones. Especially, both $R_{ph}$ and $\tau$ are strongly enhanced in the 3R bilayer BP with large interlayer distance. All these results suggest that the 3R bilayer BP structures are appealing 2D solar cell materials, and its tunable photoelectric performances provide a new strategy for future experimental design of 2D photoelectric conversion devices.

\begin{table}[t]
  \centering
  \caption{ The photon responsivity (R$_{ph}$) and the external quantum efficiency ($\tau_{eqe}$) of 2D BP based device. For comparison, the parameters of some other pervious experimental and theoretical reported 2D systems are also listed.}
    \begin{tabular}{lcccc}
    \hline
           -             & Thickness           & $R_{ph} (AW^{-1})$   & $\tau_{eqe} (\%)$ &    Ref.      \\
    \hline
    BP                              &  ML                 & 0.069            &     11.2          &      this work          \\
    3R BP$_{d=4.1\mathring{A}}$     &  ML-ML              & 0.127            &     20.7          &      this work          \\
    3R BP$_{d=3.5\mathring{A}}$     &  ML-ML              & 0.091            &     13.1          &      this work          \\
    3R BP$_{d=2.9\mathring{A}}$     &  ML-ML              & 0.113            &     14.9          &      this work          \\
    BP-BAs                          &  ML-ML              & 0.03             &     7.4           &     [\citenum{WQJMCA}]       \\
    BAs-BSb                         &  ML-ML              & 0.058            &     14.7          &     [\citenum{WQJMCA}]       \\
    KAgSe                           &  ML                 & 0.056            &     17.92         &     [\citenum{WQKAS1}]        \\
    MoS$_{2}$                       &  ML                 & 0.016            &      -            &     [\citenum{JinHao15}]   \\
    MoS$_{2}$-MoS$_{2}$             &  ML-ML              & 0.030            &      -            &     [\citenum{JinHao51}]   \\
    MoS$_{2}$-WSe$_{2}$             &  11 nm              & 0.010            &     1.5           &     [\citenum{JinHao50}]   \\
    InSe-InTe                       &  ML-ML              & 0.030            &     7.1           &     [\citenum{JinHao}]     \\
    BP-MoS$_{2}$                    & 11 nm-ML            & 0.418            &     0.3           &     [\citenum{JinHao25}]   \\
    \hline
    \end{tabular}%
  \label{tab:table2}%
\end{table}%

\section{Summary}

In summary, we carried out a first principles calculation to investigate the photovoltaic properties of 3R bilayer BX(X=P, As, Sb) and explore their potential applications as 2D XSCs. Due to the intrinsic OOP ferroelectricity induced by inversion symmetry broken, the CBM and VBM are dominated by different layers and a type-II band alignment is formed spontaneously. Lots of intriguing performances are demonstrated for these systems, including moderate band gaps, ultrahigh carrier mobilities, efficient optical absorption in visible light region and large PCE. These essential parameters of photoelectric materials can be further enhanced and a desirable red-shift of optical absorption peak is presented by increasing the interlayer distance. Excellent photoelectric conversion behaviors are also detected in the 3R bilayer BP based nanodevices. Remarkable photocurrents, photon responsivities and external quantum efficiencies are also obtained. These ideal properties reveal an appealing application of 3R bilayer BX in the photoelectric conversion and XSCs. Moveover, the efficient and adjustable photovoltaic performance driven by intrinsic OOP ferroelectricity in 3R bilayer BX provides a new strategy for designing various nanoscale ferroelectric-photovoltaic devices.

\section*{Conflicts of interest}

The authors declare no competing financial interest.

\section*{Acknowledgements}

This work was financially supported by grants from the National Natural Science Foundation of China (Grant No. 11774238), Key Program Project of National Natural Science Foundation of China (Grant No. 12034014), and Shenzhen Natural Science Foundations (Grant No. JCYJ20190808150409413).

%\bigskip
%\noindent{$^{\dag)}$ binwang@szu.edu.cn} \\
%\bigskip

%%%END OF MAIN TEXT%%%


%\balance


%%%REFERENCES%%%
\bibliography{main} %You need to replace "rsc" on this line with the name of your .bib file

\providecommand*{\mcitethebibliography}{\thebibliography}
\csname @ifundefined\endcsname{endmcitethebibliography}
{\let\endmcitethebibliography\endthebibliography}{}
\begin{mcitethebibliography}{83}
\providecommand*{\natexlab}[1]{#1}
\providecommand*{\mciteSetBstSublistMode}[1]{}
\providecommand*{\mciteSetBstMaxWidthForm}[2]{}
\providecommand*{\mciteBstWouldAddEndPuncttrue}
  {\def\EndOfBibitem{\unskip.}}
\providecommand*{\mciteBstWouldAddEndPunctfalse}
  {\let\EndOfBibitem\relax}
\providecommand*{\mciteSetBstMidEndSepPunct}[3]{}
\providecommand*{\mciteSetBstSublistLabelBeginEnd}[3]{}
\providecommand*{\EndOfBibitem}{}
\mciteSetBstSublistMode{f}
\mciteSetBstMaxWidthForm{subitem}
{(\emph{\alph{mcitesubitemcount}})}
\mciteSetBstSublistLabelBeginEnd{\mcitemaxwidthsubitemform\space}
{\relax}{\relax}

\bibitem[Haeni \emph{et~al.}(2004)Haeni, Irvin, Chang, Uecker, Reiche, Li,
  Choudhury, Tian, Hawley, Craigo,\emph{et~al.}]{OOP1}
J.~Haeni, P.~Irvin, W.~Chang, R.~Uecker, P.~Reiche, Y.~Li, S.~Choudhury,
  W.~Tian, M.~Hawley, B.~Craigo \emph{et~al.}, \emph{Nature}, 2004,
  \textbf{430}, 758--761\relax
\mciteBstWouldAddEndPuncttrue
\mciteSetBstMidEndSepPunct{\mcitedefaultmidpunct}
{\mcitedefaultendpunct}{\mcitedefaultseppunct}\relax
\EndOfBibitem
\bibitem[Valasek(1921)]{OOP2}
J.~Valasek, \emph{Physical review}, 1921, \textbf{17}, 475\relax
\mciteBstWouldAddEndPuncttrue
\mciteSetBstMidEndSepPunct{\mcitedefaultmidpunct}
{\mcitedefaultendpunct}{\mcitedefaultseppunct}\relax
\EndOfBibitem
\bibitem[Cohen(1992)]{OOP3}
R.~E. Cohen, \emph{Nature}, 1992, \textbf{358}, 136--138\relax
\mciteBstWouldAddEndPuncttrue
\mciteSetBstMidEndSepPunct{\mcitedefaultmidpunct}
{\mcitedefaultendpunct}{\mcitedefaultseppunct}\relax
\EndOfBibitem
\bibitem[Zheng \emph{et~al.}(2010)Zheng, Ni, Toh, Tan, Yao, and
  {\"O}zyilmaz]{AM26}
Y.~Zheng, G.-X. Ni, C.-T. Toh, C.-Y. Tan, K.~Yao and B.~{\"O}zyilmaz,
  \emph{Physical review letters}, 2010, \textbf{105}, 166602\relax
\mciteBstWouldAddEndPuncttrue
\mciteSetBstMidEndSepPunct{\mcitedefaultmidpunct}
{\mcitedefaultendpunct}{\mcitedefaultseppunct}\relax
\EndOfBibitem
\bibitem[Baeumer \emph{et~al.}(2015)Baeumer, Saldana-Greco, Martirez, Rappe,
  Shim, and Martin]{AM30}
C.~Baeumer, D.~Saldana-Greco, J.~M.~P. Martirez, A.~M. Rappe, M.~Shim and L.~W.
  Martin, \emph{Nature communications}, 2015, \textbf{6}, 1--9\relax
\mciteBstWouldAddEndPuncttrue
\mciteSetBstMidEndSepPunct{\mcitedefaultmidpunct}
{\mcitedefaultendpunct}{\mcitedefaultseppunct}\relax
\EndOfBibitem
\bibitem[Peng \emph{et~al.}(2014)Peng, Chen, Chen, Li, Zhao, Kou, Xiao, and
  Zhu]{OOP4}
Z.~Peng, Y.~Chen, Q.~Chen, N.~Li, X.~Zhao, C.~Kou, D.~Xiao and J.~Zhu,
  \emph{Journal of alloys and compounds}, 2014, \textbf{590}, 210--214\relax
\mciteBstWouldAddEndPuncttrue
\mciteSetBstMidEndSepPunct{\mcitedefaultmidpunct}
{\mcitedefaultendpunct}{\mcitedefaultseppunct}\relax
\EndOfBibitem
\bibitem[Bona and Smith(1975)]{OOP5}
J.~L. Bona and R.~Smith, \emph{Philosophical Transactions of the Royal Society
  of London. Series A, Mathematical and Physical Sciences}, 1975, \textbf{278},
  555--601\relax
\mciteBstWouldAddEndPuncttrue
\mciteSetBstMidEndSepPunct{\mcitedefaultmidpunct}
{\mcitedefaultendpunct}{\mcitedefaultseppunct}\relax
\EndOfBibitem
\bibitem[Moya \emph{et~al.}(2014)Moya, Kar-Narayan, and Mathur]{OOP6}
X.~Moya, S.~Kar-Narayan and N.~D. Mathur, \emph{Nature materials}, 2014,
  \textbf{13}, 439--450\relax
\mciteBstWouldAddEndPuncttrue
\mciteSetBstMidEndSepPunct{\mcitedefaultmidpunct}
{\mcitedefaultendpunct}{\mcitedefaultseppunct}\relax
\EndOfBibitem
\bibitem[Wang \emph{et~al.}(2018)Wang, Liu, Yoong, Paudel, Xiao, Guo, Lin,
  Yang, Wang, Chow,\emph{et~al.}]{OOP7}
H.~Wang, Z.~Liu, H.~Yoong, T.~R. Paudel, J.~Xiao, R.~Guo, W.~Lin, P.~Yang,
  J.~Wang, G.~Chow \emph{et~al.}, \emph{Nature communications}, 2018,
  \textbf{9}, 1--8\relax
\mciteBstWouldAddEndPuncttrue
\mciteSetBstMidEndSepPunct{\mcitedefaultmidpunct}
{\mcitedefaultendpunct}{\mcitedefaultseppunct}\relax
\EndOfBibitem
\bibitem[Junquera and Ghosez(2003)]{ACS3}
J.~Junquera and P.~Ghosez, \emph{Nature}, 2003, \textbf{422}, 506--509\relax
\mciteBstWouldAddEndPuncttrue
\mciteSetBstMidEndSepPunct{\mcitedefaultmidpunct}
{\mcitedefaultendpunct}{\mcitedefaultseppunct}\relax
\EndOfBibitem
\bibitem[Ghosez and Rabe(2000)]{ACS4}
P.~Ghosez and K.~Rabe, \emph{Applied Physics Letters}, 2000, \textbf{76},
  2767--2769\relax
\mciteBstWouldAddEndPuncttrue
\mciteSetBstMidEndSepPunct{\mcitedefaultmidpunct}
{\mcitedefaultendpunct}{\mcitedefaultseppunct}\relax
\EndOfBibitem
\bibitem[Junquera and Ghosez(2003)]{OOP10}
J.~Junquera and P.~Ghosez, \emph{Nature}, 2003, \textbf{422}, 506--509\relax
\mciteBstWouldAddEndPuncttrue
\mciteSetBstMidEndSepPunct{\mcitedefaultmidpunct}
{\mcitedefaultendpunct}{\mcitedefaultseppunct}\relax
\EndOfBibitem
\bibitem[Fong \emph{et~al.}(2004)Fong, Stephenson, Streiffer, Eastman,
  Auciello, Fuoss, and Thompson]{OOP11}
D.~D. Fong, G.~B. Stephenson, S.~K. Streiffer, J.~A. Eastman, O.~Auciello,
  P.~H. Fuoss and C.~Thompson, \emph{Science}, 2004, \textbf{304},
  1650--1653\relax
\mciteBstWouldAddEndPuncttrue
\mciteSetBstMidEndSepPunct{\mcitedefaultmidpunct}
{\mcitedefaultendpunct}{\mcitedefaultseppunct}\relax
\EndOfBibitem
\bibitem[Ding \emph{et~al.}(2017)Ding, Zhu, Wang, Gao, Xiao, Gu, Zhang, and
  Zhu]{OOP16}
W.~Ding, J.~Zhu, Z.~Wang, Y.~Gao, D.~Xiao, Y.~Gu, Z.~Zhang and W.~Zhu,
  \emph{Nature communications}, 2017, \textbf{8}, 1--8\relax
\mciteBstWouldAddEndPuncttrue
\mciteSetBstMidEndSepPunct{\mcitedefaultmidpunct}
{\mcitedefaultendpunct}{\mcitedefaultseppunct}\relax
\EndOfBibitem
\bibitem[Yuan \emph{et~al.}(2019)Yuan, Luo, Chan, Xiao, Dai, Xie, and
  Hao]{OOP14}
S.~Yuan, X.~Luo, H.~L. Chan, C.~Xiao, Y.~Dai, M.~Xie and J.~Hao, \emph{Nature
  communications}, 2019, \textbf{10}, 1--6\relax
\mciteBstWouldAddEndPuncttrue
\mciteSetBstMidEndSepPunct{\mcitedefaultmidpunct}
{\mcitedefaultendpunct}{\mcitedefaultseppunct}\relax
\EndOfBibitem
\bibitem[Fei \emph{et~al.}(2018)Fei, Zhao, Palomaki, Sun, Miller, Zhao, Yan,
  Xu, and Cobden]{OOP15}
Z.~Fei, W.~Zhao, T.~A. Palomaki, B.~Sun, M.~K. Miller, Z.~Zhao, J.~Yan, X.~Xu
  and D.~H. Cobden, \emph{Nature}, 2018, \textbf{560}, 336--339\relax
\mciteBstWouldAddEndPuncttrue
\mciteSetBstMidEndSepPunct{\mcitedefaultmidpunct}
{\mcitedefaultendpunct}{\mcitedefaultseppunct}\relax
\EndOfBibitem
\bibitem[Chandrasekaran \emph{et~al.}(2017)Chandrasekaran, Mishra, and
  Singh]{OOP18}
A.~Chandrasekaran, A.~Mishra and A.~K. Singh, \emph{Nano letters}, 2017,
  \textbf{17}, 3290--3296\relax
\mciteBstWouldAddEndPuncttrue
\mciteSetBstMidEndSepPunct{\mcitedefaultmidpunct}
{\mcitedefaultendpunct}{\mcitedefaultseppunct}\relax
\EndOfBibitem
\bibitem[Liang \emph{et~al.}(2021)Liang, Guo, Shen, Huang, Dai, and Ma]{OOP}
Y.~Liang, R.~Guo, S.~Shen, B.~Huang, Y.~Dai and Y.~Ma, \emph{Applied Physics
  Letters}, 2021, \textbf{118}, 012905\relax
\mciteBstWouldAddEndPuncttrue
\mciteSetBstMidEndSepPunct{\mcitedefaultmidpunct}
{\mcitedefaultendpunct}{\mcitedefaultseppunct}\relax
\EndOfBibitem
\bibitem[Ginley \emph{et~al.}(2008)Ginley, Green, and Collins]{LYJMCA4}
D.~Ginley, M.~A. Green and R.~Collins, \emph{Mrs Bulletin}, 2008, \textbf{33},
  355--364\relax
\mciteBstWouldAddEndPuncttrue
\mciteSetBstMidEndSepPunct{\mcitedefaultmidpunct}
{\mcitedefaultendpunct}{\mcitedefaultseppunct}\relax
\EndOfBibitem
\bibitem[Mayer \emph{et~al.}(2007)Mayer, Scully, Hardin, Rowell, and
  McGehee]{LYJMCA5}
A.~C. Mayer, S.~R. Scully, B.~E. Hardin, M.~W. Rowell and M.~D. McGehee,
  \emph{Materials today}, 2007, \textbf{10}, 28--33\relax
\mciteBstWouldAddEndPuncttrue
\mciteSetBstMidEndSepPunct{\mcitedefaultmidpunct}
{\mcitedefaultendpunct}{\mcitedefaultseppunct}\relax
\EndOfBibitem
\bibitem[Nelson(2011)]{LYJMCA6}
J.~Nelson, \emph{Materials today}, 2011, \textbf{14}, 462--470\relax
\mciteBstWouldAddEndPuncttrue
\mciteSetBstMidEndSepPunct{\mcitedefaultmidpunct}
{\mcitedefaultendpunct}{\mcitedefaultseppunct}\relax
\EndOfBibitem
\bibitem[Hong \emph{et~al.}(2014)Hong, Kim, Shi, Zhang, Jin, Sun, Tongay, Wu,
  Zhang, and Wang]{LYJMCA7}
X.~Hong, J.~Kim, S.-F. Shi, Y.~Zhang, C.~Jin, Y.~Sun, S.~Tongay, J.~Wu,
  Y.~Zhang and F.~Wang, \emph{Nature nanotechnology}, 2014, \textbf{9},
  682\relax
\mciteBstWouldAddEndPuncttrue
\mciteSetBstMidEndSepPunct{\mcitedefaultmidpunct}
{\mcitedefaultendpunct}{\mcitedefaultseppunct}\relax
\EndOfBibitem
\bibitem[Tsai \emph{et~al.}(2014)Tsai, Su, Chang, Tsai, Chen, Wu, Li, Chen, and
  He]{LYJMCA9}
M.-L. Tsai, S.-H. Su, J.-K. Chang, D.-S. Tsai, C.-H. Chen, C.-I. Wu, L.-J. Li,
  L.-J. Chen and J.-H. He, \emph{ACS nano}, 2014, \textbf{8}, 8317--8322\relax
\mciteBstWouldAddEndPuncttrue
\mciteSetBstMidEndSepPunct{\mcitedefaultmidpunct}
{\mcitedefaultendpunct}{\mcitedefaultseppunct}\relax
\EndOfBibitem
\bibitem[Britnell \emph{et~al.}(2013)Britnell, Ribeiro, Eckmann, Jalil, Belle,
  Mishchenko, Kim, Gorbachev, Georgiou, Morozov,\emph{et~al.}]{LYJMCA11}
L.~Britnell, R.~Ribeiro, A.~Eckmann, R.~Jalil, B.~Belle, A.~Mishchenko, Y.-J.
  Kim, R.~Gorbachev, T.~Georgiou, S.~Morozov \emph{et~al.}, \emph{Science},
  2013, \textbf{340}, 1311--1314\relax
\mciteBstWouldAddEndPuncttrue
\mciteSetBstMidEndSepPunct{\mcitedefaultmidpunct}
{\mcitedefaultendpunct}{\mcitedefaultseppunct}\relax
\EndOfBibitem
\bibitem[Long \emph{et~al.}(2013)Long, Dai, and Huang]{LYJMCA12}
R.~Long, Y.~Dai and B.~Huang, \emph{The Journal of Physical Chemistry Letters},
  2013, \textbf{4}, 2223--2229\relax
\mciteBstWouldAddEndPuncttrue
\mciteSetBstMidEndSepPunct{\mcitedefaultmidpunct}
{\mcitedefaultendpunct}{\mcitedefaultseppunct}\relax
\EndOfBibitem
\bibitem[Long(2013)]{LYJMCA13}
R.~Long, \emph{The journal of physical chemistry letters}, 2013, \textbf{4},
  1340--1346\relax
\mciteBstWouldAddEndPuncttrue
\mciteSetBstMidEndSepPunct{\mcitedefaultmidpunct}
{\mcitedefaultendpunct}{\mcitedefaultseppunct}\relax
\EndOfBibitem
\bibitem[Wang \emph{et~al.}(2019)Wang, Li, Liang, Wang, and Nie]{WQJMCA}
Q.~Wang, J.~Li, Y.~Liang, B.~Wang and Y.~Nie, \emph{Journal of Materials
  Chemistry A}, 2019, \textbf{7}, 10684--10695\relax
\mciteBstWouldAddEndPuncttrue
\mciteSetBstMidEndSepPunct{\mcitedefaultmidpunct}
{\mcitedefaultendpunct}{\mcitedefaultseppunct}\relax
\EndOfBibitem
\bibitem[Scharber \emph{et~al.}(2006)Scharber, M{\"u}hlbacher, Koppe, Denk,
  Waldauf, Heeger, and Brabec]{ZPJMCA13}
M.~C. Scharber, D.~M{\"u}hlbacher, M.~Koppe, P.~Denk, C.~Waldauf, A.~J. Heeger
  and C.~J. Brabec, \emph{Advanced materials}, 2006, \textbf{18},
  789--794\relax
\mciteBstWouldAddEndPuncttrue
\mciteSetBstMidEndSepPunct{\mcitedefaultmidpunct}
{\mcitedefaultendpunct}{\mcitedefaultseppunct}\relax
\EndOfBibitem
\bibitem[Liang \emph{et~al.}(2018)Liang, Dai, Ma, Ju, Wei, and Huang]{WQKAS18}
Y.~Liang, Y.~Dai, Y.~Ma, L.~Ju, W.~Wei and B.~Huang, \emph{Journal of Materials
  Chemistry A}, 2018, \textbf{6}, 2073--2080\relax
\mciteBstWouldAddEndPuncttrue
\mciteSetBstMidEndSepPunct{\mcitedefaultmidpunct}
{\mcitedefaultendpunct}{\mcitedefaultseppunct}\relax
\EndOfBibitem
\bibitem[Wang \emph{et~al.}(2018)Wang, Li, Liang, Nie, and Wang]{WQKAS1}
Q.~Wang, J.~Li, Y.~Liang, Y.~Nie and B.~Wang, \emph{ACS applied materials \&
  interfaces}, 2018, \textbf{10}, 41670--41677\relax
\mciteBstWouldAddEndPuncttrue
\mciteSetBstMidEndSepPunct{\mcitedefaultmidpunct}
{\mcitedefaultendpunct}{\mcitedefaultseppunct}\relax
\EndOfBibitem
\bibitem[Massicotte \emph{et~al.}(2016)Massicotte, Schmidt, Vialla,
  Sch{\"a}dler, Reserbat-Plantey, Watanabe, Taniguchi, Tielrooij, and
  Koppens]{JMCC23}
M.~Massicotte, P.~Schmidt, F.~Vialla, K.~G. Sch{\"a}dler, A.~Reserbat-Plantey,
  K.~Watanabe, T.~Taniguchi, K.-J. Tielrooij and F.~H. Koppens, \emph{Nature
  nanotechnology}, 2016, \textbf{11}, 42--46\relax
\mciteBstWouldAddEndPuncttrue
\mciteSetBstMidEndSepPunct{\mcitedefaultmidpunct}
{\mcitedefaultendpunct}{\mcitedefaultseppunct}\relax
\EndOfBibitem
\bibitem[Wang \emph{et~al.}(2020)Wang, Liang, Yao, Li, Wang, and Wang]{WQJMCC}
Q.~Wang, Y.~Liang, H.~Yao, J.~Li, B.~Wang and J.~Wang, \emph{Journal of
  Materials Chemistry C}, 2020, \textbf{8}, 8107--8119\relax
\mciteBstWouldAddEndPuncttrue
\mciteSetBstMidEndSepPunct{\mcitedefaultmidpunct}
{\mcitedefaultendpunct}{\mcitedefaultseppunct}\relax
\EndOfBibitem
\bibitem[Zhang \emph{et~al.}(2020)Zhang, Zhu, and Yakobson]{MoS2}
J.-J. Zhang, D.~Zhu and B.~I. Yakobson, \emph{Nano Letters}, 2020\relax
\mciteBstWouldAddEndPuncttrue
\mciteSetBstMidEndSepPunct{\mcitedefaultmidpunct}
{\mcitedefaultendpunct}{\mcitedefaultseppunct}\relax
\EndOfBibitem
\bibitem[Wang \emph{et~al.}(2015)Wang, Wang, Wang, Hu, Zhou, Guo, Huang, Sun,
  Shen, Lin,\emph{et~al.}]{MoS21}
X.~Wang, P.~Wang, J.~Wang, W.~Hu, X.~Zhou, N.~Guo, H.~Huang, S.~Sun, H.~Shen,
  T.~Lin \emph{et~al.}, \emph{Advanced materials}, 2015, \textbf{27},
  6575--6581\relax
\mciteBstWouldAddEndPuncttrue
\mciteSetBstMidEndSepPunct{\mcitedefaultmidpunct}
{\mcitedefaultendpunct}{\mcitedefaultseppunct}\relax
\EndOfBibitem
\bibitem[Liu \emph{et~al.}(2020)Liu, Pyatakov, and Ren]{VS2}
X.~Liu, A.~P. Pyatakov and W.~Ren, \emph{Physical Review Letters}, 2020,
  \textbf{125}, 247601\relax
\mciteBstWouldAddEndPuncttrue
\mciteSetBstMidEndSepPunct{\mcitedefaultmidpunct}
{\mcitedefaultendpunct}{\mcitedefaultseppunct}\relax
\EndOfBibitem
\bibitem[Liang \emph{et~al.}(2021)Liang, Mao, Dai, Kou, Huang, and Ma]{LLYY}
Y.~Liang, N.~Mao, Y.~Dai, L.~Kou, B.~Huang and Y.~Ma, \emph{arXiv preprint
  arXiv:2102.07294}, 2021\relax
\mciteBstWouldAddEndPuncttrue
\mciteSetBstMidEndSepPunct{\mcitedefaultmidpunct}
{\mcitedefaultendpunct}{\mcitedefaultseppunct}\relax
\EndOfBibitem
\bibitem[Yasuda \emph{et~al.}(2020)Yasuda, Wang, Watanabe, Taniguchi, and
  Jarillo-Herrero]{BN}
K.~Yasuda, X.~Wang, K.~Watanabe, T.~Taniguchi and P.~Jarillo-Herrero,
  \emph{arXiv preprint arXiv:2010.06600}, 2020\relax
\mciteBstWouldAddEndPuncttrue
\mciteSetBstMidEndSepPunct{\mcitedefaultmidpunct}
{\mcitedefaultendpunct}{\mcitedefaultseppunct}\relax
\EndOfBibitem
\bibitem[Xie \emph{et~al.}(2016)Xie, Zhang, Cai, Zhu, Zou, and Zeng]{BX}
M.~Xie, S.~Zhang, B.~Cai, Z.~Zhu, Y.~Zou and H.~Zeng, \emph{Nanoscale}, 2016,
  \textbf{8}, 13407--13413\relax
\mciteBstWouldAddEndPuncttrue
\mciteSetBstMidEndSepPunct{\mcitedefaultmidpunct}
{\mcitedefaultendpunct}{\mcitedefaultseppunct}\relax
\EndOfBibitem
\bibitem[Kresse and Hafner(1993)]{JMCA40}
G.~Kresse and J.~Hafner, \emph{Phys. Rev. B: Condens. Matter Mater. Phys.},
  1993, \textbf{47}, 558\relax
\mciteBstWouldAddEndPuncttrue
\mciteSetBstMidEndSepPunct{\mcitedefaultmidpunct}
{\mcitedefaultendpunct}{\mcitedefaultseppunct}\relax
\EndOfBibitem
\bibitem[Perdew \emph{et~al.}(1996)Perdew, Burke, and Ernzerhof]{JMCA41}
J.~P. Perdew, K.~Burke and M.~Ernzerhof, \emph{Phys. Rev. Lett.}, 1996,
  \textbf{77}, 3865\relax
\mciteBstWouldAddEndPuncttrue
\mciteSetBstMidEndSepPunct{\mcitedefaultmidpunct}
{\mcitedefaultendpunct}{\mcitedefaultseppunct}\relax
\EndOfBibitem
\bibitem[Heyd \emph{et~al.}(2003)Heyd, Scuseria, and Ernzerhof]{JMCA44}
J.~Heyd, G.~E. Scuseria and M.~Ernzerhof, \emph{J. Chem. Phys.}, 2003,
  \textbf{118}, 8207--8215\relax
\mciteBstWouldAddEndPuncttrue
\mciteSetBstMidEndSepPunct{\mcitedefaultmidpunct}
{\mcitedefaultendpunct}{\mcitedefaultseppunct}\relax
\EndOfBibitem
\bibitem[Kresse and Joubert(1999)]{PAW}
G.~Kresse and D.~Joubert, \emph{Physical review b}, 1999, \textbf{59},
  1758\relax
\mciteBstWouldAddEndPuncttrue
\mciteSetBstMidEndSepPunct{\mcitedefaultmidpunct}
{\mcitedefaultendpunct}{\mcitedefaultseppunct}\relax
\EndOfBibitem
\bibitem[Togo and Tanaka(2015)]{KAS36}
A.~Togo and I.~Tanaka, \emph{Scripta Materialia}, 2015, \textbf{108},
  1--5\relax
\mciteBstWouldAddEndPuncttrue
\mciteSetBstMidEndSepPunct{\mcitedefaultmidpunct}
{\mcitedefaultendpunct}{\mcitedefaultseppunct}\relax
\EndOfBibitem
\bibitem[Henkelman \emph{et~al.}(2000)Henkelman, Uberuaga, and
  J{\'o}nsson]{OOP34}
G.~Henkelman, B.~P. Uberuaga and H.~J{\'o}nsson, \emph{The Journal of chemical
  physics}, 2000, \textbf{113}, 9901--9904\relax
\mciteBstWouldAddEndPuncttrue
\mciteSetBstMidEndSepPunct{\mcitedefaultmidpunct}
{\mcitedefaultendpunct}{\mcitedefaultseppunct}\relax
\EndOfBibitem
\bibitem[Li and Wu(2017)]{OOP19}
L.~Li and M.~Wu, \emph{ACS nano}, 2017, \textbf{11}, 6382--6388\relax
\mciteBstWouldAddEndPuncttrue
\mciteSetBstMidEndSepPunct{\mcitedefaultmidpunct}
{\mcitedefaultendpunct}{\mcitedefaultseppunct}\relax
\EndOfBibitem
\bibitem[Lin \emph{et~al.}(2019)Lin, Si, Duan, Wang, and Duan]{OOP23}
Z.~Lin, C.~Si, S.~Duan, C.~Wang and W.~Duan, \emph{Physical Review B}, 2019,
  \textbf{100}, 155408\relax
\mciteBstWouldAddEndPuncttrue
\mciteSetBstMidEndSepPunct{\mcitedefaultmidpunct}
{\mcitedefaultendpunct}{\mcitedefaultseppunct}\relax
\EndOfBibitem
\bibitem[Taylor \emph{et~al.}(2001)Taylor, Guo, and Wang]{nanodcal1}
J.~Taylor, H.~Guo and J.~Wang, \emph{Phys. Rev. B}, 2001, \textbf{63},
  245407\relax
\mciteBstWouldAddEndPuncttrue
\mciteSetBstMidEndSepPunct{\mcitedefaultmidpunct}
{\mcitedefaultendpunct}{\mcitedefaultseppunct}\relax
\EndOfBibitem
\bibitem[Waldron \emph{et~al.}(2006)Waldron, Haney, Larade, MacDonald, and
  Guo]{nanodcal2}
D.~Waldron, P.~Haney, B.~Larade, A.~MacDonald and H.~Guo, \emph{Phys. Rev.
  Lett.}, 2006, \textbf{96}, 166804\relax
\mciteBstWouldAddEndPuncttrue
\mciteSetBstMidEndSepPunct{\mcitedefaultmidpunct}
{\mcitedefaultendpunct}{\mcitedefaultseppunct}\relax
\EndOfBibitem
\bibitem[Wang \emph{et~al.}(2009)Wang, Wang, and Guo]{KAS122}
B.~Wang, J.~Wang and H.~Guo, \emph{Physical Review B}, 2009, \textbf{79},
  165417\relax
\mciteBstWouldAddEndPuncttrue
\mciteSetBstMidEndSepPunct{\mcitedefaultmidpunct}
{\mcitedefaultendpunct}{\mcitedefaultseppunct}\relax
\EndOfBibitem
\bibitem[Hamann \emph{et~al.}(1979)Hamann, Schl{\"u}ter, and Chiang]{KAS51}
D.~Hamann, M.~Schl{\"u}ter and C.~Chiang, \emph{Phys. Rev. Lett.}, 1979,
  \textbf{43}, 1494\relax
\mciteBstWouldAddEndPuncttrue
\mciteSetBstMidEndSepPunct{\mcitedefaultmidpunct}
{\mcitedefaultendpunct}{\mcitedefaultseppunct}\relax
\EndOfBibitem
\bibitem[Perdew \emph{et~al.}(1996)Perdew, Burke, and Ernzerhof]{KAS52}
J.~P. Perdew, K.~Burke and M.~Ernzerhof, \emph{Phys. Rev. Lett.}, 1996,
  \textbf{77}, 3865\relax
\mciteBstWouldAddEndPuncttrue
\mciteSetBstMidEndSepPunct{\mcitedefaultmidpunct}
{\mcitedefaultendpunct}{\mcitedefaultseppunct}\relax
\EndOfBibitem
\bibitem[Qiao \emph{et~al.}(2017)Qiao, Fu, Li, Ghosen, Zeng, Stebbins, Prasad,
  and Swihart]{ACSnano}
L.~Qiao, Z.~Fu, J.~Li, J.~Ghosen, M.~Zeng, J.~Stebbins, P.~N. Prasad and M.~T.
  Swihart, \emph{ACS nano}, 2017, \textbf{11}, 6370--6381\relax
\mciteBstWouldAddEndPuncttrue
\mciteSetBstMidEndSepPunct{\mcitedefaultmidpunct}
{\mcitedefaultendpunct}{\mcitedefaultseppunct}\relax
\EndOfBibitem
\bibitem[Zhao \emph{et~al.}(2020)Zhao, Xiao, and Yao]{ZhaoSP}
P.~Zhao, C.~Xiao and W.~Yao, \emph{arXiv preprint arXiv:2011.03933}, 2020\relax
\mciteBstWouldAddEndPuncttrue
\mciteSetBstMidEndSepPunct{\mcitedefaultmidpunct}
{\mcitedefaultendpunct}{\mcitedefaultseppunct}\relax
\EndOfBibitem
\bibitem[Shirodkar and Waghmare(2014)]{ACSnano11}
S.~N. Shirodkar and U.~V. Waghmare, \emph{Physical review letters}, 2014,
  \textbf{112}, 157601\relax
\mciteBstWouldAddEndPuncttrue
\mciteSetBstMidEndSepPunct{\mcitedefaultmidpunct}
{\mcitedefaultendpunct}{\mcitedefaultseppunct}\relax
\EndOfBibitem
\bibitem[Park \emph{et~al.}(2019)Park, Yeu, Han, Hwang, and Choi]{scireport}
J.~Park, I.~W. Yeu, G.~Han, C.~S. Hwang and J.-H. Choi, \emph{Scientific
  reports}, 2019, \textbf{9}, 1--9\relax
\mciteBstWouldAddEndPuncttrue
\mciteSetBstMidEndSepPunct{\mcitedefaultmidpunct}
{\mcitedefaultendpunct}{\mcitedefaultseppunct}\relax
\EndOfBibitem
\bibitem[Jiang \emph{et~al.}(2019)Jiang, Feng, Chen, and Tang]{OOP40}
X.~Jiang, Y.~Feng, K.-Q. Chen and L.-M. Tang, \emph{Journal of Physics:
  Condensed Matter}, 2019, \textbf{32}, 105501\relax
\mciteBstWouldAddEndPuncttrue
\mciteSetBstMidEndSepPunct{\mcitedefaultmidpunct}
{\mcitedefaultendpunct}{\mcitedefaultseppunct}\relax
\EndOfBibitem
\bibitem[Kou \emph{et~al.}(2018)Kou, Fu, Ma, Yan, Liao, Du, and Chen]{OOP41}
L.~Kou, H.~Fu, Y.~Ma, B.~Yan, T.~Liao, A.~Du and C.~Chen, \emph{Physical Review
  B}, 2018, \textbf{97}, 075429\relax
\mciteBstWouldAddEndPuncttrue
\mciteSetBstMidEndSepPunct{\mcitedefaultmidpunct}
{\mcitedefaultendpunct}{\mcitedefaultseppunct}\relax
\EndOfBibitem
\bibitem[Liu \emph{et~al.}(2019)Liu, Guan, Yin, Wan, Wang, and Zhang]{OOP42}
C.~Liu, S.~Guan, H.~Yin, W.~Wan, Y.~Wang and Y.~Zhang, \emph{Applied Physics
  Letters}, 2019, \textbf{115}, 252904\relax
\mciteBstWouldAddEndPuncttrue
\mciteSetBstMidEndSepPunct{\mcitedefaultmidpunct}
{\mcitedefaultendpunct}{\mcitedefaultseppunct}\relax
\EndOfBibitem
\bibitem[Fan \emph{et~al.}(2012)Fan, Tian, and Wang]{ACSnano21}
F.-R. Fan, Z.-Q. Tian and Z.~L. Wang, \emph{Nano energy}, 2012, \textbf{1},
  328--334\relax
\mciteBstWouldAddEndPuncttrue
\mciteSetBstMidEndSepPunct{\mcitedefaultmidpunct}
{\mcitedefaultendpunct}{\mcitedefaultseppunct}\relax
\EndOfBibitem
\bibitem[Zhang \emph{et~al.}(2016)Zhang, Zhang, Chen, Jin, Deng, Tang, Zhang,
  Pan, Zhu, Yang,\emph{et~al.}]{ACSnano22}
L.~Zhang, B.~Zhang, J.~Chen, L.~Jin, W.~Deng, J.~Tang, H.~Zhang, H.~Pan,
  M.~Zhu, W.~Yang \emph{et~al.}, \emph{Advanced Materials}, 2016, \textbf{28},
  1650--1656\relax
\mciteBstWouldAddEndPuncttrue
\mciteSetBstMidEndSepPunct{\mcitedefaultmidpunct}
{\mcitedefaultendpunct}{\mcitedefaultseppunct}\relax
\EndOfBibitem
\bibitem[Zhang \emph{et~al.}(2015)Zhang, Jin, Zhang, Deng, Pan, Tang, Zhu, and
  Yang]{ACSnano23}
L.~Zhang, L.~Jin, B.~Zhang, W.~Deng, H.~Pan, J.~Tang, M.~Zhu and W.~Yang,
  \emph{Nano Energy}, 2015, \textbf{16}, 516--523\relax
\mciteBstWouldAddEndPuncttrue
\mciteSetBstMidEndSepPunct{\mcitedefaultmidpunct}
{\mcitedefaultendpunct}{\mcitedefaultseppunct}\relax
\EndOfBibitem
\bibitem[Lei \emph{et~al.}(2019)Lei, Ma, Xu, Zhang, Huang, and Dai]{jpcc121}
C.~Lei, Y.~Ma, X.~Xu, T.~Zhang, B.~Huang and Y.~Dai, \emph{The Journal of
  Physical Chemistry C}, 2019, \textbf{123}, 23089--23095\relax
\mciteBstWouldAddEndPuncttrue
\mciteSetBstMidEndSepPunct{\mcitedefaultmidpunct}
{\mcitedefaultendpunct}{\mcitedefaultseppunct}\relax
\EndOfBibitem
\bibitem[Xia \emph{et~al.}(2018)Xia, Du, Li, Li, Zhao, Wang, and Li]{PRA}
C.~Xia, J.~Du, M.~Li, X.~Li, X.~Zhao, T.~Wang and J.~Li, \emph{Physical Review
  Applied}, 2018, \textbf{10}, 054064\relax
\mciteBstWouldAddEndPuncttrue
\mciteSetBstMidEndSepPunct{\mcitedefaultmidpunct}
{\mcitedefaultendpunct}{\mcitedefaultseppunct}\relax
\EndOfBibitem
\bibitem[Cheng \emph{et~al.}(2018)Cheng, Guo, Han, Jiang, Zhang, Ahuja, Su, and
  Zhao]{cheng20182d}
K.~Cheng, Y.~Guo, N.~Han, X.~Jiang, J.~Zhang, R.~Ahuja, Y.~Su and J.~Zhao,
  \emph{Applied Physics Letters}, 2018, \textbf{112}, 143902\relax
\mciteBstWouldAddEndPuncttrue
\mciteSetBstMidEndSepPunct{\mcitedefaultmidpunct}
{\mcitedefaultendpunct}{\mcitedefaultseppunct}\relax
\EndOfBibitem
\bibitem[Bardeen and Shockley(1950)]{JMCA631}
J.~Bardeen and W.~Shockley, \emph{Physical review}, 1950, \textbf{80}, 72\relax
\mciteBstWouldAddEndPuncttrue
\mciteSetBstMidEndSepPunct{\mcitedefaultmidpunct}
{\mcitedefaultendpunct}{\mcitedefaultseppunct}\relax
\EndOfBibitem
\bibitem[Bassani \emph{et~al.}(1976)Bassani, Parravicini, Ballinger, and
  Birman]{JMCA560}
F.~Bassani, G.~P. Parravicini, R.~A. Ballinger and J.~L. Birman, \emph{Physics
  Today}, 1976, \textbf{29}, 58\relax
\mciteBstWouldAddEndPuncttrue
\mciteSetBstMidEndSepPunct{\mcitedefaultmidpunct}
{\mcitedefaultendpunct}{\mcitedefaultseppunct}\relax
\EndOfBibitem
\bibitem[Tongay \emph{et~al.}(2014)Tongay, Fan, Kang, Park, Koldemir, Suh,
  Narang, Liu, Ji, Li,\emph{et~al.}]{JMCC64}
S.~Tongay, W.~Fan, J.~Kang, J.~Park, U.~Koldemir, J.~Suh, D.~S. Narang, K.~Liu,
  J.~Ji, J.~Li \emph{et~al.}, \emph{Nano letters}, 2014, \textbf{14},
  3185--3190\relax
\mciteBstWouldAddEndPuncttrue
\mciteSetBstMidEndSepPunct{\mcitedefaultmidpunct}
{\mcitedefaultendpunct}{\mcitedefaultseppunct}\relax
\EndOfBibitem
\bibitem[Yankowitz \emph{et~al.}(2016)Yankowitz, Watanabe, Taniguchi, San-Jose,
  and LeRoy]{JMCC65}
M.~Yankowitz, K.~Watanabe, T.~Taniguchi, P.~San-Jose and B.~J. LeRoy,
  \emph{Nature communications}, 2016, \textbf{7}, 1--8\relax
\mciteBstWouldAddEndPuncttrue
\mciteSetBstMidEndSepPunct{\mcitedefaultmidpunct}
{\mcitedefaultendpunct}{\mcitedefaultseppunct}\relax
\EndOfBibitem
\bibitem[Scharber \emph{et~al.}(2006)Scharber, M{\"u}hlbacher, Koppe, Denk,
  Waldauf, Heeger, and Brabec]{JMCA60}
M.~C. Scharber, D.~M{\"u}hlbacher, M.~Koppe, P.~Denk, C.~Waldauf, A.~J. Heeger
  and C.~J. Brabec, \emph{Advanced materials}, 2006, \textbf{18},
  789--794\relax
\mciteBstWouldAddEndPuncttrue
\mciteSetBstMidEndSepPunct{\mcitedefaultmidpunct}
{\mcitedefaultendpunct}{\mcitedefaultseppunct}\relax
\EndOfBibitem
\bibitem[Zhao \emph{et~al.}(2016)Zhao, Li, Yang, Jiang, Lin, Ade, Ma, and
  Yan]{JMCA61}
J.~Zhao, Y.~Li, G.~Yang, K.~Jiang, H.~Lin, H.~Ade, W.~Ma and H.~Yan,
  \emph{Nature Energy}, 2016, \textbf{1}, 1--7\relax
\mciteBstWouldAddEndPuncttrue
\mciteSetBstMidEndSepPunct{\mcitedefaultmidpunct}
{\mcitedefaultendpunct}{\mcitedefaultseppunct}\relax
\EndOfBibitem
\bibitem[Tsai \emph{et~al.}(2014)Tsai, Su, Chang, Tsai, Chen, Wu, Li, Chen, and
  He]{JMCA62}
M.-L. Tsai, S.-H. Su, J.-K. Chang, D.-S. Tsai, C.-H. Chen, C.-I. Wu, L.-J. Li,
  L.-J. Chen and J.-H. He, \emph{ACS nano}, 2014, \textbf{8}, 8317--8322\relax
\mciteBstWouldAddEndPuncttrue
\mciteSetBstMidEndSepPunct{\mcitedefaultmidpunct}
{\mcitedefaultendpunct}{\mcitedefaultseppunct}\relax
\EndOfBibitem
\bibitem[Zhou \emph{et~al.}(2020)Zhou, Gong, Jiang, Xu, Shang, Zhang, Hu, and
  Chu]{zhoubinJMCC}
B.~Zhou, S.-J. Gong, K.~Jiang, L.~Xu, L.~Shang, J.~Zhang, Z.~Hu and J.~Chu,
  \emph{Journal of Materials Chemistry C}, 2020, \textbf{8}, 89--97\relax
\mciteBstWouldAddEndPuncttrue
\mciteSetBstMidEndSepPunct{\mcitedefaultmidpunct}
{\mcitedefaultendpunct}{\mcitedefaultseppunct}\relax
\EndOfBibitem
\bibitem[Zhang \emph{et~al.}(2014)Zhang, Gong, Chen, Liu, Zhu, Xiao, and
  Guo]{JMCA65}
L.~Zhang, K.~Gong, J.~Chen, L.~Liu, Y.~Zhu, D.~Xiao and H.~Guo, \emph{Physical
  Review B}, 2014, \textbf{90}, 195428\relax
\mciteBstWouldAddEndPuncttrue
\mciteSetBstMidEndSepPunct{\mcitedefaultmidpunct}
{\mcitedefaultendpunct}{\mcitedefaultseppunct}\relax
\EndOfBibitem
\bibitem[Xie \emph{et~al.}(2015)Xie, Zhang, Zhu, Liu, and Guo]{JMCA66}
Y.~Xie, L.~Zhang, Y.~Zhu, L.~Liu and H.~Guo, \emph{Nanotechnology}, 2015,
  \textbf{26}, 455202\relax
\mciteBstWouldAddEndPuncttrue
\mciteSetBstMidEndSepPunct{\mcitedefaultmidpunct}
{\mcitedefaultendpunct}{\mcitedefaultseppunct}\relax
\EndOfBibitem
\bibitem[Henrickson(2002)]{JMCA67}
L.~E. Henrickson, \emph{Journal of applied physics}, 2002, \textbf{91},
  6273--6281\relax
\mciteBstWouldAddEndPuncttrue
\mciteSetBstMidEndSepPunct{\mcitedefaultmidpunct}
{\mcitedefaultendpunct}{\mcitedefaultseppunct}\relax
\EndOfBibitem
\bibitem[Stewart and L{\'e}onard(2004)]{JMCA68}
D.~Stewart and F.~L{\'e}onard, \emph{Physical review letters}, 2004,
  \textbf{93}, 107401\relax
\mciteBstWouldAddEndPuncttrue
\mciteSetBstMidEndSepPunct{\mcitedefaultmidpunct}
{\mcitedefaultendpunct}{\mcitedefaultseppunct}\relax
\EndOfBibitem
\bibitem[Koppens and Mueller(2014)]{xxx11}
F.~Koppens and T.~Mueller, \emph{Photodetectors based on graphene, other
  two-dimensional materials and hybrid systems. Nat. Nanotechnol}, 2014,
  \textbf{9}, 780--793\relax
\mciteBstWouldAddEndPuncttrue
\mciteSetBstMidEndSepPunct{\mcitedefaultmidpunct}
{\mcitedefaultendpunct}{\mcitedefaultseppunct}\relax
\EndOfBibitem
\bibitem[Wang \emph{et~al.}(2015)Wang, Wang, Xu, Wang, Wang, Huang, Yin, and
  He]{xxx22}
F.~Wang, Z.~Wang, K.~Xu, F.~Wang, Q.~Wang, Y.~Huang, L.~Yin and J.~He,
  \emph{Nano letters}, 2015, \textbf{15}, 7558--7566\relax
\mciteBstWouldAddEndPuncttrue
\mciteSetBstMidEndSepPunct{\mcitedefaultmidpunct}
{\mcitedefaultendpunct}{\mcitedefaultseppunct}\relax
\EndOfBibitem
\bibitem[Jin \emph{et~al.}(2016)Jin, Li, Wang, Yu, Wan, Xu, Dai, Wei, and
  Guo]{JinHao15}
H.~Jin, J.~Li, B.~Wang, Y.~Yu, L.~Wan, F.~Xu, Y.~Dai, Y.~Wei and H.~Guo,
  \emph{Journal of Materials Chemistry C}, 2016, \textbf{4}, 11253--11260\relax
\mciteBstWouldAddEndPuncttrue
\mciteSetBstMidEndSepPunct{\mcitedefaultmidpunct}
{\mcitedefaultendpunct}{\mcitedefaultseppunct}\relax
\EndOfBibitem
\bibitem[Li \emph{et~al.}(2015)Li, Lee, Qu, Liu, Ryu, Seabaugh, and
  Yoo]{JinHao51}
H.-M. Li, D.~Lee, D.~Qu, X.~Liu, J.~Ryu, A.~Seabaugh and W.~J. Yoo,
  \emph{Nature communications}, 2015, \textbf{6}, 1--9\relax
\mciteBstWouldAddEndPuncttrue
\mciteSetBstMidEndSepPunct{\mcitedefaultmidpunct}
{\mcitedefaultendpunct}{\mcitedefaultseppunct}\relax
\EndOfBibitem
\bibitem[Furchi \emph{et~al.}(2014)Furchi, Pospischil, Libisch, Burgdorfer, and
  Mueller]{JinHao50}
M.~M. Furchi, A.~Pospischil, F.~Libisch, J.~Burgdorfer and T.~Mueller,
  \emph{Nano letters}, 2014, \textbf{14}, 4785--4791\relax
\mciteBstWouldAddEndPuncttrue
\mciteSetBstMidEndSepPunct{\mcitedefaultmidpunct}
{\mcitedefaultendpunct}{\mcitedefaultseppunct}\relax
\EndOfBibitem
\bibitem[Jin \emph{et~al.}(2016)Jin, Li, Wang, Yu, Wan, Xu, Dai, Wei, and
  Guo]{JinHao}
H.~Jin, J.~Li, B.~Wang, Y.~Yu, L.~Wan, F.~Xu, Y.~Dai, Y.~Wei and H.~Guo,
  \emph{Journal of Materials Chemistry C}, 2016, \textbf{4}, 11253--11260\relax
\mciteBstWouldAddEndPuncttrue
\mciteSetBstMidEndSepPunct{\mcitedefaultmidpunct}
{\mcitedefaultendpunct}{\mcitedefaultseppunct}\relax
\EndOfBibitem
\bibitem[Deng \emph{et~al.}(2014)Deng, Luo, Conrad, Liu, Gong, Najmaei, Ajayan,
  Lou, Xu, and Ye]{JinHao25}
Y.~Deng, Z.~Luo, N.~J. Conrad, H.~Liu, Y.~Gong, S.~Najmaei, P.~M. Ajayan,
  J.~Lou, X.~Xu and P.~D. Ye, \emph{ACS nano}, 2014, \textbf{8},
  8292--8299\relax
\mciteBstWouldAddEndPuncttrue
\mciteSetBstMidEndSepPunct{\mcitedefaultmidpunct}
{\mcitedefaultendpunct}{\mcitedefaultseppunct}\relax
\EndOfBibitem
\end{mcitethebibliography}
\bibliographystyle{rsc} %the RSC's .bst file

\end{document}